\documentstyle[preprint,aps,pra]{revtex}

\input epsf
\tightenlines 
\begin{document}
\title{Floquet Scattering and Classical-Quantum correspondence in
strong time periodic fields}
\author{Agapi Emmanouilidou, L.E. Reichl}
\address{Center for Studies in Statistical Mechanics and Complex Systems,}
\address{The University of Texas at Austin, Austin, Texas, 78212}
\date{\today}
\maketitle

\begin{center}
{\bf{Abstract}}
\end{center}

We study the scattering of an electron from a one dimensional inverted Gaussian atomic 
potential in the presence of strong time periodic electric fields. Using Floquet theory,
 we construct the Floquet Scattering matrix in the Kramers-Henneberger frame.
 We compute the transmission coefficients as a function of electron incident 
energy
 and find that they display asymmetric Fano resonances due to the electron interaction 
with the driving field. We find that the Fano resonances are associated with zero-pole
pairs of the Floquet Scattering matrix in the complex energy plane.
 Another way we ``probe'' the complex spectrum of the system is by computing
the Wigner-Smith delay times.
 Finally we find
 that the eigenphases of the Floquet Scattering matrix undergo a number
of ``avoided crossings'' as a function of electron Floquet energy, and this
number
 increases with increasing strength of the driving field. These 
``avoided crossings'' appear to be quantum manifestations
of the destruction of the constants of motion and the onset of chaos in classical phase space.

\vspace{1cm}

\section{Introduction}
The development of ultra-high intensity lasers has led to the study
of atoms in external time periodic electric fields that are comparable
in strength to the electric fields produced by the atomic nucleus. One of the 
most 
interesting phenomena observed in these time periodic systems is the
stabilization with increasing laser intensity
that was predicted theoretically \cite{ion1,ion2,ion3} and has been verified
experimentally \cite{experim1,experim2}.     
 
 In previous studies, one dimensional atomic potentials
have been used to predict several
phenomena in the theory of laser-atom interactions at high laser
intensities. Many of these studies were carried out in the context of 
Floquet theory formulated in the Kramers-Henneberger (K-H) frame of 
reference \cite{Kramers,Henneberger}, which oscillates with a free electron in 
the time periodic field. Gavrila and Kaminski \cite{ion1} developed a 
nonperturbative method to
study electron scattering in the presence of strong time-periodic electric fields. Using three dimensional models, Dimou and Faisal \cite{Dimou} as well as 
Collins and Csanak \cite{Dimou} have studied resonances in laser assisted 
scattering. 
 Bhatt, Piraux and Burnett \cite{Bhatt} in their work on electron 
scattering from a polarization potential in the presence of strong
monochromatic light, argued the appearance of new light-induced quasibound 
states (resonances) as the field strength is increased.
 The same phenomenon was later also observed by Bardsley 
and Comella \cite{Bardsley}
and Yao and Chu \cite{Yao} who used the complex coordinate scaling 
transformation to compute the complex quasibound states   
in their study of photodetachment from a one-dimensional Gaussian potential.
 In addition, the same atomic potential was used by Marinescu and  
 Gavrila \cite{Marinescu} to compare the predictions of the full Floquet theory
with those of the high-frequency Floquet theory (HFFT), using  resonance
(Siegert) boundary conditions. (The HFFT theory is a version
of the Floquet approach adapted to treat the high frequency limit.) Recently,
 Timberlake and Reichl \cite{Timberlake2}, using the inverted Gaussian potential, studied
the phase-space picture of resonance creation and they showed that
the light-induced quasibound states are scarred on unstable periodic orbits of 
the classical motion.

In this paper, we study the scattering of an electron from a short range atomic 
potential in the presence of a strong time 
periodic electric field. The atomic potential we consider is the one-dimensional 
inverted Gaussian potential, a model that has already offered
interesting insights into different aspects of the laser-atom interactions
\cite{Bardsley,Yao,Marinescu}. Our goal is to construct the 
Floquet Scattering matrix ($\mbox{\boldmath$S$}$-matrix) using the full 
Floquet theory formulated in the 
K-H frame for a strongly driven atomic system. A Floquet Scattering 
matrix has been constructed by Li and 
Reichl \cite{Reichl1} for 
periodically driven mesoscopic systems. The Floquet $\mbox{\boldmath$S$}$-matrix 
connects the outgoing propagating modes to the incoming propagating modes and is 
a unitary matrix which conserves probability. We construct the Floquet 
$\mbox{\boldmath$S$}$-matrix in the K-H frame where we can define asymptotic 
states.

In section II, we construct the Floquet $\mbox{\boldmath$S$}$-matrix in the K-H frame.
 In section III.A we compute the transmission coefficients and the poles of the 
Floquet $\mbox{\boldmath$S$}$-matrix
and find the quasibound states of the atomic system.
 In section III.B we compute the Wigner-Smith
delay times of the scattered electron as a function of the electron incident energy 
 and show that the Wigner-Smith delay times of the scattered electron due to the
presence of the quasibound states are of the same order of magnitude as the lifetimes 
obtained from the poles of the Floquet $\mbox{\boldmath$S$}$-matrix. 
 We finish, in section III.C, with a quite interesting observation. When plotting
the eigenphases of the unitary Floquet $\mbox{\boldmath$S$}$-matrix as a function of the 
electron's Floquet energy we find that at certain energies the eigenphases 
undergo ``avoided crossings'' that change the eigenphases character completely. 
 We find that the number of ``avoided crossings'' increases with increasing 
strength of the time periodic electric field. The ``avoided crossings'' observed 
as the strength of the driving field is increased appear to be quantum 
manifestations of the destruction of 
the KAM surfaces and the onset of chaos in the classical phase space.  

\section{The Floquet  $\protect\mbox{\protect\boldmath$S$}$-matrix}

\subsection{The model} 
We study the scattering of an electron in the presence of a strong
electric field and a short-range atomic potential. The electric field 
 $E(t)=E_{0}\sin(\omega t)$ ($T=\frac{2\pi}{\omega}$ is the period of the
field) is 
treated within the dipole approximation as a monochromatic infinite plane wave 
linearly polarized along the direction of the incident electron.
The Schr\mbox{\"{o}}dinger equation, in one space dimension $x$, that describes the 
dynamics of the system is in atomic units (a.u.)  
\begin{equation}
\label{eq:Schrod}
i\frac{\partial \Psi(x,t)}{\partial t}=
\left({\frac{1}{2}(-i\frac{\partial}{\partial x}-q A(t))^2+V(x)}\right)\Psi(x,t),
\end{equation}
where V(x) is the inverted Gaussian potential:
\begin{equation}
\label{eq:pot}
V(x)=-V_{0}e^{-(x/\delta)^2},
\end{equation}
and $q$ is the particle charge which for the electron is $q=-1$ a.u.. 
The electric field is, $E(t)=-\frac{\partial A(t)}{\partial t}$, where $A(t)$ is the 
vector potential and is given by 
\begin{equation}
\label{eq:A(t)}
A(t)=\frac{E_{0}}{\omega}\cos(\omega t).
\end{equation} 
We use atomic units $(e=\hbar=m=1)$ throughout this paper, except when otherwise
indicated. 

 To construct the Floquet $\mbox{\boldmath$S$}$-matrix of the system, we transform to the 
K-H frame \cite{Kramers,Henneberger}. In the K-H frame there are well 
defined asymptotic regions and the boundary conditions
are expressed in terms of free electron waves. To obtain 
the wave function in the K-H frame we introduce the unitary 
transformation \cite{Kramers,Henneberger}    
\begin{equation}
\label{eq:Wavef}
\Phi(x,t)=U_{1}U_{2}\Psi(x,t),
\end{equation}  
where 
\begin{equation}
\label{eq:U12}
\begin{array}{ccc}
U_{1}=e^{\frac{i q^{2}}{2}\int_{-\infty}^{t} A^{2}(t')dt'}
 &\mbox{and}&
U_{2}=e^{-q\int_{-\infty}^{t} A(t')\frac{\partial}{\partial x}dt'}\end{array}.
\end{equation}
$U_{1}$ is a phase transformation to remove the $A^{2}$ term from 
Eq.(\ref{eq:Schrod}) while $U_{2}$ is a space translation transformation to the K-H frame. 
 In the K-H frame, the wave function satisfies the following Schr\mbox{\"{o}}dinger equation
\begin{equation}
\label{eq:Schrod2}
i\frac{\partial \Phi(x,t)}{\partial t}=\left (-\frac{1}{2}
\frac{\partial^{2}}{\partial x^{2}}+V(x+a(t))\right)\Phi(x,t),
\end{equation}
where $a(t)$ is the classical displacement of a free electron from its center of
oscillation in the time periodic field $E(t)$, and is given by
\begin{equation}
\label{eq:a(t)}
\begin{array}{ccc}
a(t)=-q\int_{-\infty}^{t}A(t')dt'=a_{0}\sin(\omega t) &\mbox{with} & 
a_{0}=\frac{-q E_{0}}{\omega^2}\end{array}.
\end{equation}

\begin{figure}
\begin{centering}
\leavevmode
\epsfxsize=0.5\linewidth
\epsfbox{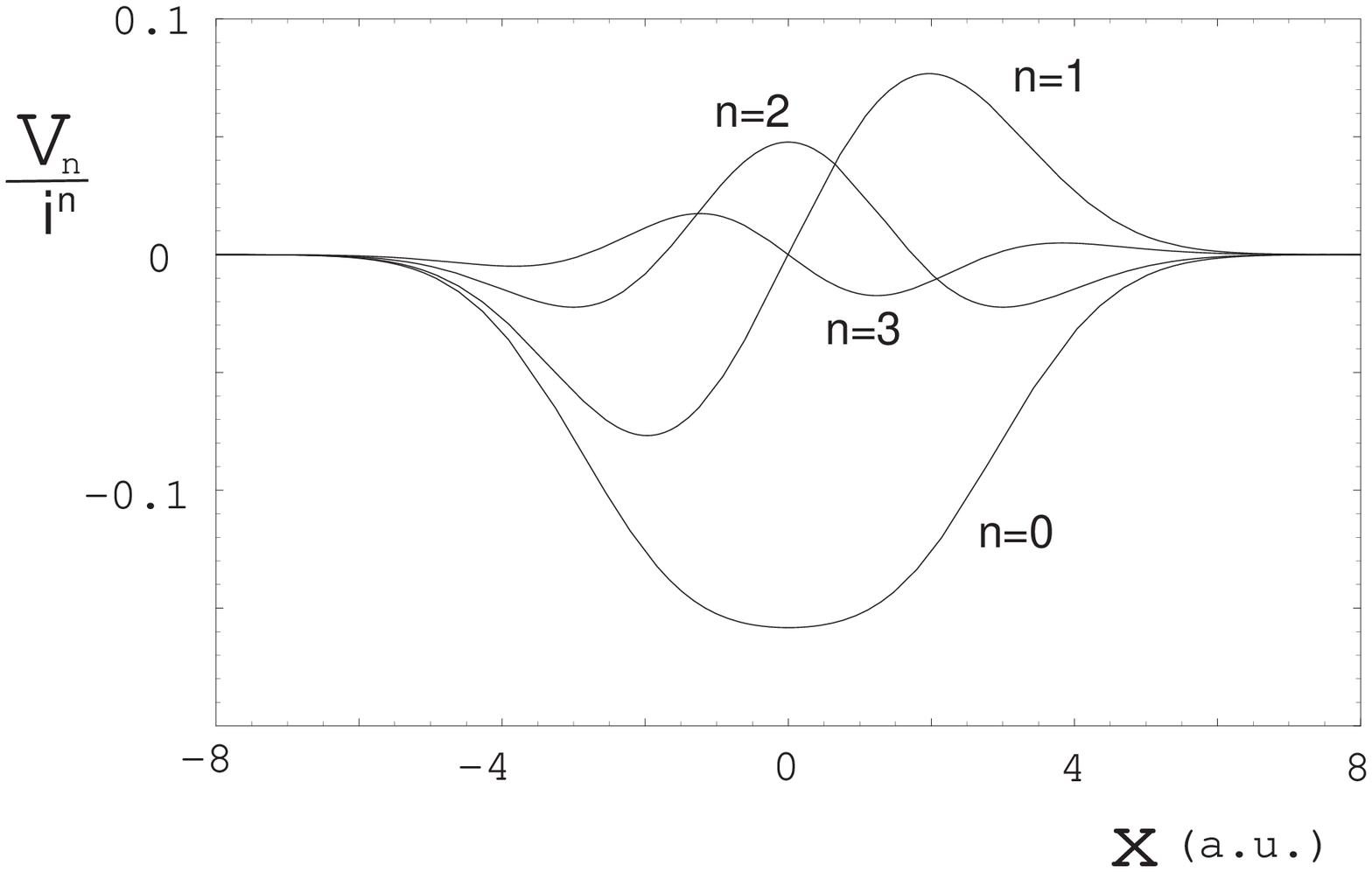}
\caption{}
\label{fig:Fourier}
\end{centering}
\end{figure}

In Eq.(\ref{eq:Schrod2}), the potential is a periodic function of time,
 that is $V(x+a(t))=V(x+a(t+T))$. Thus, according to the Floquet theorem
\cite{Floquet}, Eq.(\ref{eq:Schrod2}) has solutions of the form
\begin{equation}
\label{eq:wavef}
\Phi_{\mathcal{E}}(x,t)=e^{-i {\mathcal{E}} t}\phi_{\mathcal{E}}(x,t),
\end{equation}
where ${\mathcal{E}}$ is the Floquet energy, ${\mathcal{E}}\in [0,\omega)$, and 
$\phi_{\mathcal{E}}(x,t)$ is a periodic function of time, $\phi_{\mathcal{E}}(x,t)=\phi_{\mathcal{E}}(x,t+T)$. 
 Taking the Fourier expansion of $\phi_{\mathcal{E}}(x,t)$ we obtain
 
\begin{equation}
\label{eq:waveff}
\Phi_{\mathcal{E}}(x,t)=e^{-i {\mathcal{E}} t}\sum_{n=-\infty}^{+\infty} \phi_{n}(x) e^{-i n \omega t},
\end{equation}
where $n$ indicates the Floquet channel. Note that $\phi_{n}(x)$ is also
${\mathcal{E}}$ dependent but we omit the ${\mathcal{E}}$
subscript to simplify notation. The energy $E$ of an incident electron in the
K-H frame is related to the Floquet energy through the expression $E={\mathcal{E}}+n\omega$.
 Next, we Fourier analyze the potential
\begin{equation}
\label{eq:potf}
V(x+a(t))=\sum_{n=-\infty}^{+\infty}V_{n}(a_{0};x)e^{-inwt},  
\end{equation} 
where the Fourier components for the inverted Gaussian potential, 
 Eq.(\ref{eq:pot}), can be written as
\begin{equation}
\label{eq:potff}
V_{n}(a_{0};x)=\frac{1}{2\pi}\int_{0}^{2\pi}V(x+a(t))
e^{i n \omega t}d(\omega t)=-V_{0}\frac{i^{n}}{\pi}\int_{0}^{\pi}\cos(n\omega t)
e^{-(x+a_{0}\cos(\omega t))^{2}/\delta^{2}}d(\omega t), 
\end{equation}
see Fig.(\ref{fig:Fourier}).
 To be able to construct the Floquet $\mbox{\boldmath$S$}$-matrix, the Fourier components 
$V_{n}(a_{0};x)$ must be smooth functions in the one space dimension $x$ in 
the K-H frame. 
 This is indeed the case for the inverted Gaussian 
potential, Eq.(\ref{eq:potff}). Note that in the K-H frame the potential 
oscillates back and forth along the $x$-axis (laterally) with the period of the
external field.

\begin{figure}
\begin{centering}
\leavevmode
\epsfxsize=0.5\linewidth
\epsfbox{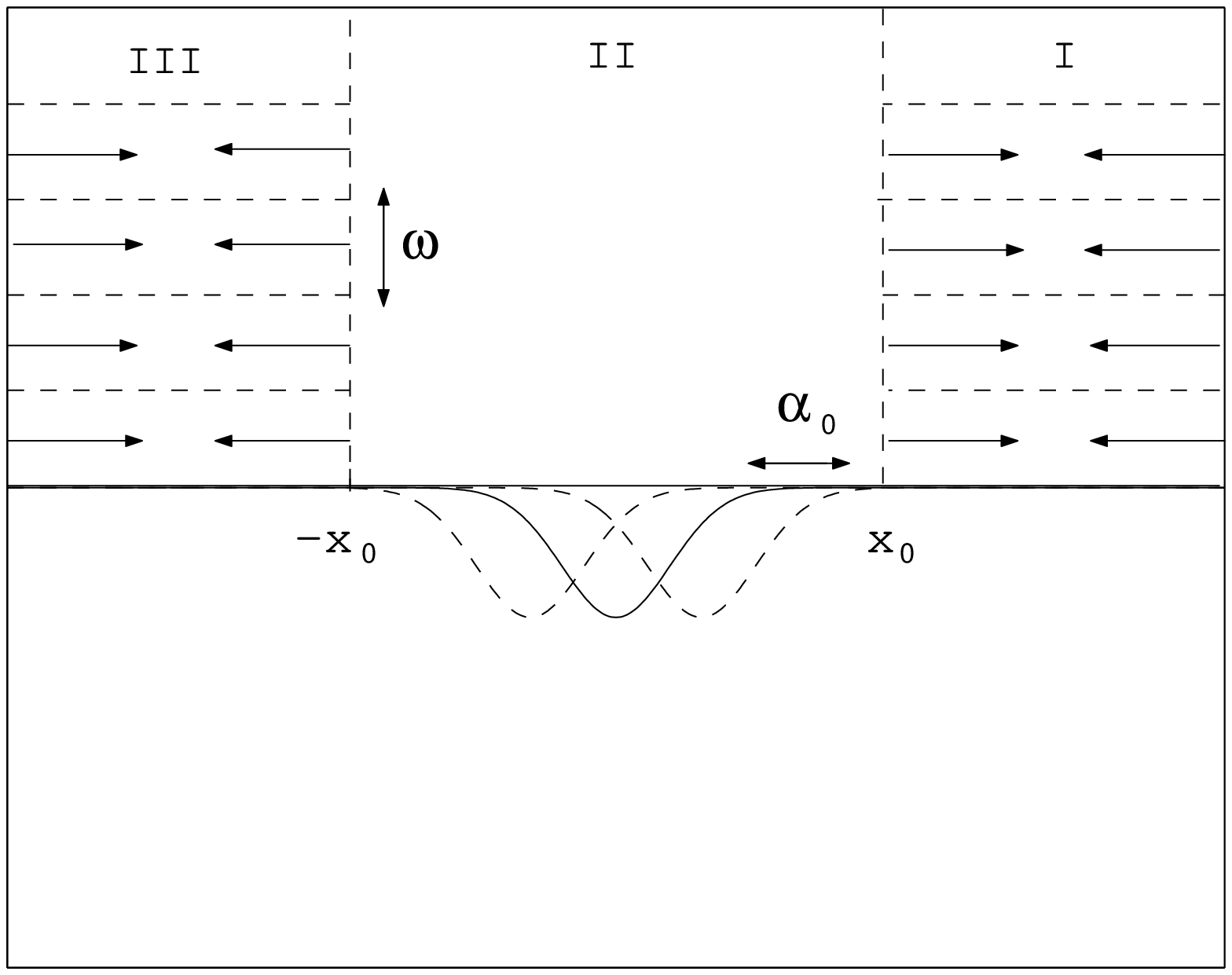}
\caption{}
\label{fig:potential}
\end{centering}
\end{figure}

From Eq.(\ref{eq:potff}), we see that the components $V_{n}(\alpha_{0};x)$
of the atomic potential in the limit $x\rightarrow \pm \infty$ tend to zero
faster than $1/x$ and we can thus divide the one space dimension $x$ in three
regions: the asymptotic regions I, $x\in [x_{0},\infty)$, and III, 
 $x\in (-\infty, -x_{0}]$, where the potential 
is asymptotically zero; and the scattering region II, $x\in [-x_{0},x_{0}]$,
where the potential $V(x+\alpha(t))$ is not zero, see 
Fig.(\ref{fig:potential}). In the rest of this paper, for brevity, we refer to the potential in asymptotic regions I and III as being zero instead of asymptotically zero.
 The choice of $x_{0}$ depends on the value of the parameter $\alpha_{0}$. 
 The larger $\alpha_{0}$ is, the further out we have to define 
the asymptotic regions I and III.

\subsection{Floquet solution in the scattering region II}

 Substituting Eqs.(\ref{eq:waveff}) and (\ref{eq:potf})
into Eq.(\ref{eq:Schrod2}) we obtain an infinite system of coupled differential
equations \cite{ion1} for the Floquet components $\phi_{n}(x)$
\begin{equation}
\label{eq:coupled}
-\frac{1}{2}\frac{d^{2}}{dx^{2}}\phi_{n}(x)+[V_{0}(\alpha_{0};x)-({\mathcal{E}}+n\omega)]
\phi_{n}(x)+\sum_{\stackrel{l=-\infty}{l\ne n}}^{+\infty}
V_{n-l}(\alpha_{0};x)\phi_{l}(x)=0.
\end{equation}
Next, we truncate to a finite number of Floquet channels and take
$n_{e}$ and $n_{p}$ to be the lower and upper limit of the Floquet channels
considered. That is, $n=-n_{e},...,0,...,n_{p}$ and the total number of
Floquet channels is given by $N=n_{e}+n_{p}+1$.   
 After truncating, Eq.(\ref{eq:coupled}) can be cast in the following 
matrix form,

\begin{equation}
\label{eq:trunsys}
\mbox{\boldmath $I$}\frac{d^2}{dx^2}\mbox{\boldmath $\phi$}^{II}(x)=
\mbox{\boldmath $M$}(x)\mbox{\boldmath $\phi$}^{II}(x),
\end{equation}
where $\mbox{\boldmath $I$}$ is the unit $N\times N$ matrix, 
 $\mbox{\boldmath $\phi$}^{II}(x)$ is the $N\times 1$ matrix
with elements $\mbox{\boldmath $\phi$}^{II}_{n}(x)=\phi_{n}(x)$ 
and $\mbox{\boldmath$M$}(x)$ is an $N\times N$ matrix with elements
\begin{equation}
\label{eq:energy}
\mbox{\boldmath$M$}_{n,l}(x)=2(V_{n-l}(\alpha_{0};x)-
\delta_{n,l}({\mathcal{E}}+n\omega)),
\end{equation}
where $\delta_{n,l}$ is the Kronecker delta and $n,l=-n_{e},...,0,...,n_{p}$.
 The general solution of the second order $N$ coupled differential equations, 
 Eq.(\ref{eq:trunsys}), can be written as a linear combination of $2N$ linearly
independent columns $\mbox{\boldmath$\chi$}_{j}(x)$, with $j=1,...,2N$, as 
follows 
\begin{eqnarray}
\label{eq:sys1}
\mbox{\boldmath $\phi$}^{II}(x)&=&
c_{1}\mbox{\boldmath$\chi$}_{1}(x)+
c_{2}\mbox{\boldmath$\chi$}_{2}(x)+\cdots
+c_{N}\mbox{\boldmath$\chi$}_{N}(x)+
d_{1}\mbox{\boldmath$\chi$}_{N+1}(x)+
d_{2}\mbox{\boldmath$\chi$}_{N+2}(x)+\cdots
+d_{N}\mbox{\boldmath$\chi$}_{2N}(x) \nonumber\\
&\equiv&\mbox{\boldmath$X$}^{(1)}(x)\mbox{\boldmath$C$}+
\mbox{\boldmath$X$}^{(2)}(x)\mbox{\boldmath $D$}
\end{eqnarray}
where $\mbox{\boldmath$X$}^{(1)}(x)$ and 
$\mbox{\boldmath$X$}^{(2)}(x)$ are $N\times N$ matrices whose elements
are functions of the one space dimension $x$ and $\mbox{\boldmath $C$}$,
 $\mbox{\boldmath $D$}$ are constant $N\times 1$ matrices. Each
of the linearly independent columns $\mbox{\boldmath$\chi$}_{j}(x)$
satisfies Eq.(\ref{eq:trunsys})

\begin{equation}
\label{eq:sys2}
\mbox{\boldmath$I$}\frac{d^2}{dx^2}\mbox{\boldmath $\chi$}_{j}(x)=
\mbox{\boldmath$M$}(x)\mbox{\boldmath $\chi$}_{j}(x)\Rightarrow
\frac{d^2}{dx^2}\chi_{n,j}(x)=\sum_{l=-n_{e}}^{n_{p}}\mbox{\boldmath$M$}_{n,l}(x)\chi_{l,j}(x),
\end{equation}  
where $n=-n_{e},...,0,...,n_{p}$ and $j=1,...,2N$. The functions $\chi_{n,j}(x)$
can be found analytically if the matrix elements of 
$\mbox{\boldmath$M$}(x)$ are constant.   
From Eq.(\ref{eq:sys1}), it follows
that every channel function
$\phi_{n}^{II}(x)$ can be written as a linear combination of $2N$ functions 
$\chi_{n,j}(x)$ and thus, the wavefunction in the scattering region II is 
given by
\begin{equation}
\label{eq:solutII}
\Phi^{II}_{\mathcal{E}}(x,t)=\sum_{n=-n_{e}}^{n_{p}}\sum_{m=1}^{N}(\chi_{n,m}(x)c_{m}+
\chi_{n,N+m}(x)d_{m})e^{-i {\mathcal{E}} t} e^{-i n \omega t}.
\end{equation}  

\subsection{Floquet solution in the asymptotic regions} 

In the asymptotic regions I and III the potential $V(x+\alpha(t))$ is zero. Thus, we 
can consider as our boundary conditions a superposition of incoming and outgoing
free electron waves in the $N$ truncated Floquet channels that are incident from
both sides of the scattering region:    

\begin{equation}
\label{eq:solutI}
\Phi^{I}_{\mathcal{E}}(x,t)=\sum_{n=-n_{e}}^{n_{p}}\phi_{n}^{I}(x)e^{-i {\mathcal{E}} t} e^{-i n \omega t}=
\sum_{n=-n_{e}}^{n_{p}}b_{n}^{out} \frac{e^{i k_{n} x}}{\sqrt{k_{n}}}
e^{-i {\mathcal{E}} t} e^{-i n \omega t} +\sum_{n=-n_{e}}^{n_{p}}b_{n}^{in} \frac{e^{-i k_{n} x}}
{\sqrt{k_{n}}} e^{-i {\mathcal{E}} t} e^{-i n \omega t},
\end{equation}

\begin{equation}
\label{eq:solutIII}
\Phi^{III}_{\mathcal{E}}(x,t)=\sum_{n=-n_{e}}^{n_{p}}\phi_{n}^{III}(x)e^{-i {\mathcal{E}} t} e^{-i n \omega t}=
\sum_{n=-n_{e}}^{n_{p}}a_{n}^{out} \frac{e^{-i k_{n} x}}{\sqrt{k_{n}}}
e^{-i {\mathcal{E}} t} e^{-i n \omega t} +\sum_{n=-n_{e}}^{n_{p}}a_{n}^{in} \frac{e^{i k_{n} x}}
{\sqrt{k_{n}}} e^{-i {\mathcal{E}} t} e^{-i n \omega t},
\end{equation}
where $b_{n}^{in}$, $a_{n}^{in}$ and $b_{n}^{out}$, $a_{n}^{out}$ are the
probability amplitudes of the incoming and outgoing electron waves, respectively,
 that are
incident in the nth Floquet channel with energy $E={\mathcal{E}}+n\omega$, see
Fig.(\ref{fig:potential}). 
 Propagating modes
are incident on the Floquet channels $n=0,...,n_{p}$ and have wavevectors 
$k_{n}=\sqrt{2({\mathcal{E}}+n\omega)}$, while evanescent modes occupy the 
Floquet channels $n=-n_{e},...,-1$ and have imaginary wavevectors $k_{n}=
i\sqrt{2|{\mathcal{E}}+n\omega|}$. The current density of the evanescent modes 
is zero. We note that the terms propagating/evanescent modes 
correspond to what some authors refer to as open/closed channels, respectively. For the Floquet $\mbox{\boldmath$S$}$-matrix
to be unitary we need to normalize the current density of the propagating modes.
 To do so, we have introduced the constants $1/\sqrt{k_{n}}$ in the
wavefunction in Eqs.(\ref{eq:solutI}) and (\ref{eq:solutIII}). To simplify notation
in Eqs.(\ref{eq:solutI}) and (\ref{eq:solutIII}), we introduce the 
constants $1/\sqrt{k_{n}}$ for the evanescent modes as well, 
 even though they have zero current density.

It is important to note, once again, that the reason we choose to work in the K-H
frame is that in this frame we can define asymptotic regions where the
potential is zero and thus, the Floquet channels are not coupled, in contrast with
the scattering region, as we have already shown. The existence of the asymptotic 
regions guarantees that
probability is conserved in the truncated number of Floquet channels and thus the
Floquet $\mbox{\boldmath$S$}$-matrix is a unitary matrix.

\subsection{Floquet $\mbox{\boldmath$S$}$-matrix}

The Floquet $\mbox{\boldmath$S$}$-matrix connects the outgoing propagating 
modes with the incoming propagating modes, and in this section we show how to
construct it. As we show in what follows, the Floquet $\mbox{\boldmath$S$}$-matrix connects channels with
energies that differ by an integer multiple of $\omega$, while in the usual
time-independent scattering theory the 
$\mbox{\boldmath$S$}$-matrix connects channels with the same energy. The reason
is that the Floquet $\mbox{\boldmath$S$}$-matrix describes a time-dependent 
process and thus the energy of the incident electron is not conserved. However,
 because the Hamiltonian is time periodic, according to Floquet theory 
\cite{Floquet}, the 
Floquet energy $\mathcal{E}$ defined modulo $\omega$ is a conserved 
quantity.

 The wavefunction and its first spatial derivative must be continuous at the 
boundaries of the asymptotic regions $\pm x_{0}$. At $x=x_{0}$ these conditions
lead to

\begin{equation}
\label{eq:contI}
b_{n}^{out} \frac{e^{i k_{n} x_{0}}}{\sqrt{k_{n}}}+b_{n}^{in}\frac{e^{-i k_{n} x_{0}}}{\sqrt{k_{n}}}=\sum_{m=1}^{N}(\chi_{n,m}(x_{0})c_{m}+\chi_{n,N+m}(x_{0})d_{m}),
\end{equation}
\begin{equation}
\label{eq:contdI}
ik_{n}b_{n}^{out} \frac{e^{i k_{n} x_{0}}}{\sqrt{k_{n}}}-ik_{n}b_{n}^{in}\frac{e^{-i k_{n} x_{0}}}{\sqrt{k_{n}}}=\sum_{m=1}^{N}(\chi'_{n,m}(x_{0})c_{m}+\chi'_{n,N+m}(x_{0})d_{m}),
\end{equation}
where $\chi'_{n,m}(x)=\frac{d \chi_{n,m}(x)}{dx}$ and  
$\chi'_{n,N+m}(x)=\frac{d \chi_{n,N+m}(x)}{dx}$, while at $x=-x_{0}$ they lead to

\begin{equation}
\label{eq:contII}
a_{n}^{out} \frac{e^{i k_{n} x_{0}}}{\sqrt{k_{n}}}+a_{n}^{in}\frac{e^{-i k_{n} x_{0}}}{\sqrt{k_{n}}}=\sum_{m=1}^{N}(\chi_{n,m}(-x_{0})c_{m}+\chi_{n,N+m}(-x_{0})d_{m}),
\end{equation}
\begin{equation}
\label{eq:contdII}
-ik_{n}a_{n}^{out} \frac{e^{i k_{n} x_{0}}}{\sqrt{k_{n}}}+ik_{n}a_{n}^{in}\frac{e^{-i k_{n} x_{0}}}{\sqrt{k_{n}}}=\sum_{m=1}^{N}(\chi'_{n,m}(-x_{0})c_{m}+\chi'_{n,N+m}(-x_{0})d_{m}).
\end{equation}

Due to the connection conditions Eqs.(\ref{eq:contI}), (\ref{eq:contdI}), (\ref{eq:contII}) and(\ref{eq:contdII}) only $2N$ out of the $6N$ coefficients
are arbitrary and we choose those to be the incoming probability amplitudes
$a^{in}$ and $b^{in}$. In Eqs.(\ref{eq:contI}), (\ref{eq:contdI}), (\ref{eq:contII}) and(\ref{eq:contdII}), the probability amplitudes $a^{in}$ and $b^{in}$ of the
evanescent modes are zero because of the unbounded character of the exponentials
they multiply in the asymptotic regions I and III. That is, $b_{n}^{in}=a_{n}^{in}=0$, for $n=-n_{e},...,-1$. 
 We now introduce the $N\times N$ matrices
\begin{equation}
\label{eq:matrix1}
\begin{array}{lll}
(\mbox{\boldmath $K$}_{\pm})_{n,l}=e^{\pm i k_{n}x_{0}}\delta_{n,l}, &
(\mbox{\boldmath $K$}'_{\pm})_{n,l}=
\pm ik_{n}e^{\pm i k_{n}x_{0}}\delta_{n,l}, 
&\mbox{$n,l=-n_{e},...,0,...,n_{p},$}\end{array}
\end{equation}

\begin{equation}
\label{eq:matrix3}
\mbox{\boldmath $J$}_{n,l}=\left \{\begin{array}{ll}
               0, & \mbox{if $n\ne l$ and if $n=l=-n_{e},...,-1,$}\\
               1, & \mbox{if $n=l=0,...,n_{p}$} 
                             \end{array}\right.
\end{equation}
and
\begin{equation}
\label{eq:matrix3a}
\mbox{\boldmath $\mathcal{N}$}_{n,l}=\begin{array}{ccc}
 \frac{1}{\sqrt{k_{n}}}\delta_{n,l}, & & \mbox{$n,l=-n_{e},...,0,...,n_{p}.$} \end{array}
\end{equation}
Also $\mbox{\boldmath $X$}^{(1)}_{\pm}\equiv
\mbox{\boldmath $X$}^{(1)}(\pm x_{0})$, $\mbox{\boldmath $X$}^{(2)}_{\pm}\equiv
\mbox{\boldmath $X$}^{(2)}(\pm x_{0})$, ${\mbox{\boldmath $X$}^{(1)}}^{'}_{\pm}\equiv
{\mbox{\boldmath $X$}^{(1)}}^{'}(\pm x_{0})$, ${\mbox{\boldmath $X$}^{(2)}}^{'}_{\pm}\equiv
{\mbox{\boldmath $X$}^{(2)}}^{'}(\pm x_{0})$,
where ${\mbox{\boldmath $X$}^{(1)}}^{'}(x)$ and ${\mbox{\boldmath $X$}^{(2)}}^{'}(x)$ are
the derivatives of $\mbox{\boldmath $X$}^{(1)}(x)$ and $\mbox{\boldmath $X$}^{(2)}(x)$
with respect to the one space dimension $x$. We also introduce the 
$N\times 1$ matrices $\mbox{\boldmath$A$}_{n}^{out}=a_{n}^{out}$, $\mbox{\boldmath$B$}_{n}^{out}=b_{n}^{out}$, $\mbox{\boldmath$A$}_{n}^{in}=a_{n}^{in}$, 
 $\mbox{\boldmath$B$}_{n}^{in}=b_{n}^{in}$. 
 Next, we write Eqs.(\ref{eq:contI}), (\ref{eq:contdI}), (\ref{eq:contII}) and
(\ref{eq:contdII}) in matrix form as follows
\begin{equation}
\label{eq:1}
\mbox{\boldmath $\mathcal{N}$}\mbox{\boldmath $K$}_{+}
\mbox{\boldmath $B$}^{out}+ \mbox{\boldmath $\mathcal{N}$}\mbox{\boldmath $K$}_{-}\mbox{\boldmath $J$}
\mbox{\boldmath $B$}^{in}= \mbox{\boldmath $X$}^{(1)}_{+}
\mbox{\boldmath $C$}+\mbox{\boldmath $X$}^{(2)}_{+}
\mbox{\boldmath $D$}
\end{equation}

\begin{equation}
\label{eq:2}
\mbox{\boldmath $\mathcal{N}$}\mbox{\boldmath $K$}'_{+}
\mbox{\boldmath $B$}^{out}+ \mbox{\boldmath $\mathcal{N}$}\mbox{\boldmath $K$}'_{-}\mbox{\boldmath $J$}
\mbox{\boldmath $B$}^{in}= {\mbox{\boldmath $X$}^{(1)}}^{'}_{+}
\mbox{\boldmath $C$}+{\mbox{\boldmath $X$}^{(2)}}^{'}_{+}
\mbox{\boldmath $D$}
\end{equation}

\begin{equation}
\label{eq:3}
\mbox{\boldmath $\mathcal{N}$}\mbox{\boldmath $K$}_{+}
\mbox{\boldmath $A$}^{out}+ \mbox{\boldmath $\mathcal{N}$}\mbox{\boldmath $K$}_{-}\mbox{\boldmath $J$}
\mbox{\boldmath $A$}^{in}= \mbox{\boldmath $X$}^{(1)}_{-}
\mbox{\boldmath $C$}+\mbox{\boldmath $X$}^{(2)}_{-}
\mbox{\boldmath $D$}
\end{equation}

\begin{equation}
\label{eq:4}
-\mbox{\boldmath $\mathcal{N}$}\mbox{\boldmath $K$}'_{+}
\mbox{\boldmath $A$}^{out}- \mbox{\boldmath $\mathcal{N}$}\mbox{\boldmath $K$}'_{-}\mbox{\boldmath $J$}
\mbox{\boldmath $A$}^{in}= {\mbox{\boldmath $X$}^{(1)}}^{'}_{-}
\mbox{\boldmath $C$}+{\mbox{\boldmath $X$}^{(2)}}^{'}_{-}
\mbox{\boldmath $D$}
\end{equation}
After algebra given in Appendix A we find the Floquet $\mbox{\boldmath$S$}$-matrix,
 that connects the outgoing probability amplitudes of the propagating modes
to the incoming probability amplitudes of the propagating modes, to be  

\begin{eqnarray}
\label{eq:13}
\left(\begin{array}{c}
                   \mbox{\boldmath $A$}_{p}^{out}\\
                    \mbox{\boldmath $B$}_{p}^{out}\end{array}\right)&=&
                  \left(\begin{array}{cc}
\mbox{\boldmath $\mathcal{N}$}_{pp}^{-1}\mbox{\boldmath $r$}_{pp}'
\mbox{\boldmath $\mathcal{N}$}_{pp}& \mbox{\boldmath $\mathcal{N}$}_{pp}^{-1}\mbox{\boldmath $t$}_{pp}
\mbox{\boldmath $\mathcal{N}$}_{pp}\\
\mbox{\boldmath $\mathcal{N}$}_{pp}^{-1}\mbox{\boldmath $t$}_{pp}'
							\mbox{\boldmath $\mathcal{N}$}_{pp}& \mbox{\boldmath $\mathcal{N}$}_{pp}^{-1}\mbox{\boldmath $r$}_{pp}
\mbox{\boldmath $\mathcal{N}$}_{pp}\end{array}\right)
\left(\begin{array}{c}
           \mbox{\boldmath $A$}_{p}^{in}\\
                    \mbox{\boldmath $B$}_{p}^{in}\end{array}\right)\nonumber\\
&\equiv&\left(\begin{array}{cc}
              \mbox{\boldmath $R$}'&\mbox{\boldmath $T$}\\
              \mbox{\boldmath $T$}'&\mbox{\boldmath $R$}\end{array}\right)
\left(\begin{array}{c}
           \mbox{\boldmath $A$}_{p}^{in}\\
                    \mbox{\boldmath $B$}_{p}^{in}\end{array}\right)
\equiv\mbox{\boldmath $S$}\left(\begin{array}{c}
           \mbox{\boldmath $A$}_{p}^{in}\\
                    \mbox{\boldmath $B$}_{p}^{in}\end{array}\right),
\end{eqnarray}
where the $n_{p}+1\times n_{p}+1$ ($n_{p}+1$
is the number of the propagating modes) matrices $\mbox{\boldmath $r$}'_{pp}$, $\mbox{\boldmath $r$}_{pp}$,
 $\mbox{\boldmath $t$}'_{pp}$ and $\mbox{\boldmath $t$}_{pp}$ defined in
Eqs.(\ref{eq:10}) and (\ref{eq:11}) of Appendix A connect propagating 
modes to propagating modes and contain the evanescent mode effect as is shown
in Appendix A.
 Also the $n_{p}+1\times 1$ matrices $\mbox{\boldmath $A$}^{in}_{p}$
 $\mbox{\boldmath $A$}^{out}_{p}$, $\mbox{\boldmath $B$}^{in}_{p}$
and $\mbox{\boldmath $B$}^{out}_{p}$ have elements the amplitudes of the propagating modes
and are defined in Eq.(\ref{eq:11b}) of Appendix A, and the 
$n_{p}+1\times n_{p}+1$ matrix  
$\mbox{\boldmath $\mathcal{N}$}_{pp}$ has elements the normalization constants
of the propagating modes and is defined in Eq.(\ref{eq:11b}) of Appendix A.   
The matrices $\mbox{\boldmath $R$}'$, $\mbox{\boldmath $R$}$, 
 $\mbox{\boldmath $T$}'$ and $\mbox{\boldmath $T$}$ have dimensions
$n_{p}+1\times n_{p}+1$ and their elements are given in terms of the elements
of the $\mbox{\boldmath $r$}'_{pp}$, $\mbox{\boldmath $r$}_{pp}$, 
 $\mbox{\boldmath $t$}'_{pp}$ and $\mbox{\boldmath $t$}_{pp}$ matrices as follows 
\begin{equation}
\begin{array}{cccc}
R'_{n',n}=\sqrt{\frac{k_{n'}}{k_{n}}}(\mbox{\boldmath$r$}'_{pp})_{n',n},
&R_{n',n}=\sqrt{\frac{k_{n'}}{k_{n}}}(\mbox{\boldmath$r$}_{pp})_{n',n},
&T'_{n',n}=\sqrt{\frac{k_{n'}}{k_{n}}}(\mbox{\boldmath$t$}'_{pp})_{n',n},
&T_{n',n}=\sqrt{\frac{k_{n'}}{k_{n}}}(\mbox{\boldmath$t$}_{pp})_{n',n}\end{array},
\end{equation}
with $n',n=0,...,n_{p}$.

In Appendix B we show how to obtain numerically the matrices 
$\mbox{\boldmath $r$}_{pp}$ and $\mbox{\boldmath $t$}_{pp}$. 
 The Floquet $\mbox{\boldmath $S$}$-matrix
has dimensions $2(n_{p}+1)\times 2(n_{p}+1)$, see Eq.(\ref{eq:13}), and is 
determined by the reflection and transmission amplitudes,
 $R'_{n',n}$, $R_{n',n}$, $T'_{n',n}$ and $T_{n',n}$, of the
propagating modes.  
 The elements $|R_{n',n}|^{2}$ and $|T_{n',n}|^{2}$
 are the 
reflection and transmission coefficients respectively for an electron wave 
incident on the propagating channel $n$ from the right that gets scattered to
the propagating channel $n'$, while the elements 
$|R'_{n',n}|^{2}$ and $|T'_{n',n}|^{2}$ are the 
reflection and transmission coefficients respectively for an electron wave 
incident on the propagating channel $n$ from the left that gets scattered to
the propagating channel $n'$.

In this section we have shown how to construct the Floquet 
$\mbox{\boldmath$S$}$-matrix in the K-H frame.
 The reason we work in the K-H frame is that we can define asymptotic
regions where the wavefunction is a superposition of free electron waves.
 That guarantees that the truncated Floquet $\mbox {\boldmath$S$}$-matrix
is a unitary matrix, that is, the following condition is satisfied
    
\begin{equation}
\label{eq:probab}
\sum_{n'=0}^{n_{p}}[|R_{n',n}|^2+|T_{n',n}|^2]=1,  
\end{equation}
for every incident propagating mode $n=0,...,n_{p}$. The above 
condition is a statement of conservation of probability. 
 Also, the Floquet $\mbox{\boldmath $S$}$-matrix we construct in
the K-H frame is isospectral with the corresponding matrix in the Lab frame
since a unitary transformation is used to transform from the Lab to the 
K-H frame, see section II.A. Finally, the criterion we use to successfully
truncate to $N$ Floquet channels is that
 an electron wave incident on the last propagating Floquet channel
$n=n_{p}$ is not affected by the scattering potential. That is,
 the transmission coefficient $|T_{n_{p},n_{p}}|^{2}$ should be equal to one
as a function of electron incident energy $E$ ( $E={\mathcal{E}}+n_{p}\omega$) as we discuss 
in more detail in section III.B.

\subsection{Symmetries of the Floquet $\mbox{\boldmath $S$}$-matrix}

The Hamiltonian of the scattering model we consider in the K-H frame,
 Eq.(\ref{eq:Schrod2}), is invariant under the transformation
$x\rightarrow -x$ and $t\rightarrow t+T/2$, which is known as Generalized
Parity. Thus, $H(x,t)=H(-x,t+T/2)$ and therefore $\Phi(-x,t+T/2)$ is also a solution
of Eq.(\ref {eq:Schrod2}).
 Applying the transformation $x\rightarrow -x$ and $t\rightarrow t+T/2$
to Eqs.(\ref{eq:solutI}) and (\ref{eq:solutIII}) it is easy to show that
the Floquet $\mbox{\boldmath $S$}$-matrix has the following symmetry:
\begin{equation}
\label{eq:symmetry}
\begin{array}{cccc}R'_{n',n}=R_{n',n}(-1)^{n'-n}, & T'_{n',n}=T_{n',n}(-1)^{n'-n},&
\mbox{with}& n',n=0,...,n_{p}\end{array}.
\end{equation}
Thus, if we know the reflection/transmission amplitudes, $R_{n',n}/T_{n',n}$ 
for electron waves incident from the right using Eqs.(\ref{eq:symmetry}) we can 
find the reflection/transmission amplitudes, $R'_{n',n}/T'_{n',n}$ for 
electron waves incident from the left and vise versa.   

\section{Results}
In this section, the calculations are performed with the values
$V_{0}=0.27035$ a.u. and $\delta=2$ a.u. assigned to the parameters of the inverted 
Gaussian potential. For these parameters the Gaussian 
potential supports only one bound state of energy $E_{b}=-0.1327$ a.u. in 
the field-free case. The parameters $V_{0}$ and $\delta$ were chosen so as 
to describe the behavior of
a one-dimensional model negative chlorine ion, $Cl^{-}$, in the presence of a 
laser field, and are the same as considered in \cite{Yao,Marinescu,Fearnside}.
 The frequency of the time periodic field is taken constant and equal
to $\omega=0.236$ a.u. for all our calculations. For these values of the parameters $V_{0}$,
 $\delta$ and $\omega$ the inverted Gaussian potential has been shown to exhibit stabilization \cite{Yao,Marinescu}.

\subsection{Transmission resonances}

In this section, we compute the transmission coefficient ${\mathit{T_{n}}}$ and the total 
transmission coefficient $\mathit{T_{tot,n}}$ as a function of the electron incident energy $E$,
 where
\begin{equation}
\label{eq:tran}
\begin{array}{ccccc}
{\mathit{T_{n}}}(E)=|T_{n,n}|^{2}, & & {\mathit{T_{tot,n}}}(E)=\sum_{n'=0}^{n_{p}}|T_{n',n}|^{2},
 & & \mbox{$E\in [n\omega, (n+1)\omega)$}\end{array}
\end{equation}
with $n=0,1$. Thus, we consider an electron wave incident from the right with energy $E\in[0,2\omega)$
and compute the transmission coefficients.   
 Keeping the frequency of the time periodic field constant, $\omega=0.236$ a.u., and
varying the strength of the driving field, $a_{0}$, we plot the transmission coefficients 
$\mathit{T_{n}}$ and $\mathit{T_{tot,n}}$ in Figs.(\ref{fig:T05}), (\ref{fig:T225}) 
and (\ref{fig:T525}) for $\alpha_{0}$ equal to $0.5$, $2.25$ and $5.25$, respectively. The frequency of the driving field $\omega=0.236$ a.u. is chosen so that it is 
larger than the binding energy of the ground state $|E_{b}|=|-0.1327|$ a.u. in the 
field-free case.
 The Floquet channels we retain to obtain the 
numerical results presented in sections III.A and III.B are $n=-6,...,0,...,6$ for 
$\alpha_{0}=0.5$, $n=-12,...,0,...,12$ for $\alpha_{0}=2.25$ and
 $n=-19,...,0,...,19$ for $\alpha_{0}=5.25$, for reasons we discuss in detail at the end 
of section III.B. 

The transmission coefficients $\mathit{T_{n}}$ and $\mathit{T_{tot,n}}$ display sharp 
asymmetric resonances, as a function of electron incident energy $E$, that 
involve a dip or a transmission peak/dip as is shown
in Figs.(\ref{fig:T05}), (\ref{fig:T225}) and (\ref{fig:T525}). These asymmetric
resonances are due to the interaction of the incident electron wave with the 
laterally oscillating potential in the K-H frame and are the so-called
Fano \cite{Fano,Tekman} type transmission resonances that are known to occur when a
bound state is coupled to a continuum of states. This is indeed the case for the scattering
model we consider, where the bound state of the inverted Gaussian potential is 
coupled to a continuum of states through the time periodic electric field. Note in
Figs.(\ref{fig:T05}), (\ref{fig:T225}) and (\ref{fig:T525}) that
the difference between the transmission coefficient $\mathit{T_{n}}$ and 
$\mathit{T_{tot,n}}$ becomes more prominent with increasing $\alpha_{0}$. The reason is 
that as $\alpha_{0}$ is increased more Floquet channels interfere with the incident electron
wave and significantly contribute to the total transmission coefficient. A 
comparison of Fig.(\ref{fig:T05}) with Figs.(\ref{fig:T225}) and (\ref{fig:T525}) 
reveals that as the driving field is increased the higher
order resonances, for $E >\omega$, become stronger.
   
 We now focus on the transmission coefficient $\mathit{T_{n}}$ and discuss how 
it ``probes'' the quasibound states of the system. 
 For $\alpha_{0}=0.5$, see Fig.(3a) the system has only one Fano transmission
resonance which for small amplitude of the driving field is associated with 
the $n=-1$ localized Floquet evanescent mode which has its origin in the bound state 
of the undriven system. When the strength of the driving field
is increased, $\mathit{T_{n}}$ has a second Fano transmission resonance 
at a higher incident energy, see Figs.(4a) and (5a). This second resonance appears 
for $a_{0}>1$, as was shown in \cite{Yao,Marinescu}, and it is thus a field induced
resonance. 

 The Fano-resonances, which are indicated by a dip or a transmission peak/dip in the 
coefficient $\mathit{T_{n}}$, correspond to quasibound states of the system that 
show up as poles of the Floquet $\mbox{\boldmath$S$}$-matrix in the complex 
energy plane. In what follows, we compute the poles of $\mathit{T_{n}}$
in the complex energy plane. Other elements of the Floquet $\mbox{\boldmath$S$}$-matrix 
have poles as well. As was noted in \cite{pairs} the asymmetric Fano 
line shape in $\mathit{T_{n}}$ is associated
with zero-pole pairs when plotting $\mathit{T_{n}}$ in the complex energy plane.  
 By a zero-pole pair we mean that every transmission zero of $\mathit{T_{n}}$ along the
real energy axis is associated with a pole of $\mathit{T_{n}}$ on the lower half 
complex energy plane due to the unitarity of the Floquet 
$\mbox{\boldmath$S$}$-matrix \cite{pairs}. For $\alpha_{0}=0.5$ there is only one 
zero-pole pair associated with the single transmission resonance seen in 
$\mathit{T_{n}}$, while for $\alpha_{0}=2.25$ there is a zero-pole pair for
each of the two resonances, see Figs.(4a) and (\ref{fig:225cont}). For small 
strengths of the driving field, $\alpha_{0}=0.5$, the location of the 
pole on the lower half complex energy plane and of the zero on the real energy axis is the same, 
while there is a small difference for stronger fields, $\alpha_{0}=2.25$. 
 That is why, we can only approximately determine the real part of the quasibound states 
from the transmission zeros. From the poles in the complex energy plane, we find the 
real part of the quasibound states 
to be $Re(E_{1})=0.106$, for $\alpha_{0}=0.5$, and $Re(E_{1})=0.145$, $Re(E_{2})=0.226$,
 for $\alpha_{0}=2.25$. 

The lifetime, $\tau_{L}$ of the quasibound states is determined from the imaginary part
of the complex energy, $Im(E)$, where the pole is found. Then
\begin{equation}
\label{eq:timep}
\tau_{L}=\frac{1}{ \Gamma},
\end{equation}
where $\Gamma=2 Im(E)$ is the ionization rate. For the inverted Gaussian potential
 it has been found that with increasing strength of the driving field the 
ionization rate decreases in an oscillatory manner \cite{Yao,Marinescu}. 
In Figs.(\ref{fig:energyr}) and (\ref{fig:energyi}) we show how the real and imaginary 
part of the quasibound state energies change as a function of $\alpha_{0}$, for $\alpha_{0}$
ranging from $0$ to $6$ a.u.. The incident particle can emit a photon and drop to a 
localized Floquet evanescent state. It is in this sense that in Fig.(\ref{fig:energyr})
we plot the real part of the quasibound state minus a photon energy and obtain results in
agreement with those obtained in \cite{Yao,Marinescu}. 

\subsection{Wigner-Smith delay times}

Wigner's \cite{Wigner} one dimensional analysis on time delay in a quantum 
mechanical scattering problem was generalized to multi-channel scattering
by Smith \cite{Smith} who introduced the Hermitian
matrix 
\begin{equation}
\hat{Q}=i\hat{S}
\frac{d \hat{S}^{\dagger}}{dE}
\label{eq:Qmatrix}
\end{equation}
  and interpreted its diagonal elements $\mbox{\boldmath$Q$}_{nn}$
 as the average delay experienced by a particle incident
on the nth channel ($\mbox{\boldmath$S$}$ is the unitary scattering matrix). 

In what follows, we compute the diagonal elements of the Wigner-Smith delay
matrix $\mbox{\boldmath$Q$}$ for the system currently under consideration.
 We first obtain the eigenvalues and eigenvectors of
the Floquet $\mbox{\boldmath $S$}$-matrix, see Eq.(\ref{eq:13}), which is a 
$2(n_{p}+1)\times 2(n_{p}+1)$ matrix when the system is truncated to $n_{p}+1$
propagating modes. The $2(n_{p}+1)$ eigenvalues of the unitary Floquet 
$\mbox{\boldmath $S$}$-matrix have unit magnitude and can thus be cast in the form 
$e^{i\theta_{i}({\mathcal{E}})}$,
 where $i=1,...,2(n_{p}+1)$ and $\theta_{i}(\mathcal{E})$ is the ith eigenphase
as a function of the Floquet energy $\mathcal{E}$. The eigenvector 
corresponding to the ith eigenvalue $e^{i\theta_{i}}$ is denoted by 
$|\theta_{i}>$. We note that the transmission coefficients $\mathit{T_{n}}$ and 
$\mathit{T_{tot,n}}$ (see section III.A) as well as the Wigner-Smith delay times 
$\tau_{ws}^{n}$, defined in what follows, are a function of incident electron energy
$E\in[0, (n_{p}+1)\omega)$. However, the eigenphases $\theta_{i}$ and the eigenvectors 
$|\theta_{i}>$ of the Floquet $\mbox{\boldmath $S$}$-matrix are a function of the 
Floquet energy ${\mathcal{E}} \in[0,\omega)$ in the sense that if the Floquet energy
is defined as the incident electron energy at a higher propagating channel one finds
the exact same eigenvalues and eigenvectors.
 The Floquet $\mbox{\boldmath$S$}$-matrix 
can be written as
\begin{equation}
\hat{S}=\sum_{i=1}^{2(n_{p}+1)}|\theta_{i}> e^{i\theta_{i}}<\theta_{i}|.
\label{eq:SSmatrix}
\end{equation}   
 Each of the eigenvectors $|\theta_{i}>$ can be expanded in terms of the propagating
free electron waves $|k_{n}>$ that we have used to
construct the Floquet $\mbox{\boldmath$S$}$-matrix, where $<x|k_{n}>=e^{i k_{n} x}$. That is,
 $|\theta_{i}>=\sum_{n=0}^{2n_{p}+1}p_{n,i}|k_{n}>$ where $p_{n,i}=<k_{n}|\theta_{i}>$
and $P_{n,i}=|p_{n,i}|^{2}$ is the occupation probability of 
the $|\theta_{i}>$ eigenvector on the $|k_{n}>$ propagating channel and $n=0,...,n_{p}$
for the right propagating modes and $n=n_{p}+1,...,2n_{p}+1$ for the left propagating
modes. The occupation probability of the $|\theta_{i}>$ eigenvector
on a mode incident from the right is the same with that of the corresponding mode
incident from the left, that is, $P_{n,i}=P_{n+n_{p}+1,i}$ for $n=0,...,n_{p}$.
 Finally, for each eigenvector $|\theta_{i}>$ the total occupation probability
is normalized to 1, $\sum_{n=0}^{2n_{p}+1}P_{n,i}=1$, where modes incident from the 
right and the left are taken into account. For modes only incident from the right
the normalization for the occupation probability takes the form
$\sum_{n=0}^{n_{p}}P_{n,i}=0.5$. 
From Eqs. (\ref{eq:Qmatrix}) and (\ref{eq:SSmatrix}) and the fact that the eigenvectors $|\theta_{i}>$ form a complete set (the Floquet $\mbox{\boldmath$S$}$-matrix is unitary) one can show that
\begin{equation}
<k_{n}|\hat{Q}|k_{n}>\equiv \tau_{ws}^{n}=\sum_{i=1}^{2(n_{p}+1)}\frac{d \theta_{i}}
{d {\mathcal{E}}}<k_{n}|\theta_{i}><\theta_{i}|k_{n}>\Rightarrow \tau_{ws}^{n}=\sum_{i=1}^{2(n_{p}+1)}P_{n,i}\frac{d \theta_{i}}{d{\mathcal{E}}}.
\label{eq:tws}
\end{equation} 
The Wigner-Smith delay times $\tau_{ws}^{n}$ are the average times an electron
incident on the nth channel with energy $E\in[n\omega, (n+1)\omega)$ is delayed due to its interaction with the laterally oscillating time periodic potential in the K-H frame. The Wigner-Smith delay times for propagating modes incident from the right are the same with those incident from the left, 
 that is, $\tau_{ws}^{n}=\tau_{ws}^{n+n_{p}+1}$ for $n=0,...,n_{p}$. 

In Figs.(\ref{fig:wig05}), (\ref{fig:wig225}) and (\ref{fig:wig525}) we plot the Wigner-Smith delay
times for $\alpha_{0}$ equal to $0.5$, $2.25$ and $5.25$, respectively, for
modes incident from the right. In Table I we compare
the Wigner-Smith delay times $\tau_{ws}^{n}$ obtained from the Floquet $\mbox{\boldmath$S$}$-matrix 
with the lifetime $\tau_{L}$ obtained from the poles of the ${\it{T_{n}}}$ transmission coefficient in 
the complex energy plane and find them 
to be of the same order of magnitude \cite{Na}.
 At the transmission
resonances, the incident electron wave gets trapped by the oscillating
potential, populating the quasibound states of the system. The delay 
of the incident electron wave at the transmission resonances shows up as
peaks when plotting the Wigner-Smith delay times as a function of the electron 
incident energy.
   Note that
as $\alpha_{0}$ increases from $2.25$ to $5.25$ the Wigner-Smith delay time of the 
1st 
quasibound state increases, evidence of stabilization. Another interesting
observation is that for small incident energy $E$ the electron has positive Wigner-Smith 
delay times, for $\alpha_{0}=0.5$, but the electron has negative Wigner-Smith delay times
for strong driving fields $\alpha_{0}=2.25$ and $5.25$. These last can arise
physically either from reflection of the incident electron before it enters the
scattering region or from its acceleration and swift passage through the negative 
potential \cite{Smith}. In addition, from Figs.(\ref{fig:wig05}), (\ref{fig:wig225}) and (\ref{fig:wig525}) we see, as expected, that when the electron is incident on 
higher Floquet channels it delays less and less until for very high energies
it is not affected by the potential and the delay time is zero.  

Now, let us briefly comment on the truncation error 
of our numerical calculations. The number of Floquet channels $N$ was chosen
for each value of $\alpha_{0}$ so that the error due to truncation remains
small. The truncation error for the elements $R_{n',n}/T_{n',n}$, where 
$n',n=0,...,n_{p}$, is smaller for $n=0$  and it increases as 
$n$ approaches $n_{p}$, where $n_{p}$ is the last propagating mode and thus 
the mode with the larger electron incident energy. Thus, the truncation error
for the transmission coefficients $\mathit{T_{n}}$ and $\mathit{T_{tot,n}}$ computed 
in section III.A is smaller than the error for the Wigner-Smith delay times computed 
in this section.
 An estimate of the truncation error is given by $1-|T_{n_{p},n_{p}}|^{2}$ as 
a function of incident electron energy. In all our calculations the truncation error is kept in the order of $10^{-4}$ for $a_{0}=0.5, 2.25$ and $10^{-3}$ for
$a_{0}=5.25$ so that our results are reliable.  As we increase $\alpha_{0}$ we need to consider
a larger number of Floquet channels $N$ to maintain a small truncation error in our numerical
calculations making it computationally challenging to compute the Wigner-Smith delay times for 
large values of $\alpha_{0}$. 

\subsection{Classical-Quantum Correspondence}
 
 When we plot the eigenphases of the Floquet $\mbox{\boldmath$S$}$-matrix as a function of 
the electron Floquet energy ${\mathcal{E}}$ we notice that the eigenphases undergo an increasing number of ``avoided 
crossings'' with increasing strength $\alpha_{0}$ of the driving field, see 
Figs.(\ref{fig:phases05}), (\ref{fig:phases125}), and (\ref{fig:phases225}). As we show
in what follows, we believe that these ``avoided crossings'' are a quantum manifestation
of chaos in the classical phase space.  
    
Let us explain what we mean by the term ``avoided crossing'' in terms
of the occupation probabilities, $P_{n,i}$, (defined in section III.B) of the $|\theta_{i}>$ eigenvector on the $|k_{n}>$ propagating channel. In what follows we consider the occupation probabilities only for modes incident from the 
right, that is, $n=0,...,n_{p}$. 
In Fig.(\ref{fig:phases05}) the eigenphases $\theta_{1}$ and $\theta_{2}$ undergo
a repulsion when the Floquet energy is equal to the transmission 
resonance, ${\mathcal{E}}=0.106$, for $\alpha_{0}=0.5$. For very small values of $\alpha_{0}$ the eigenphases cross each other without repelling. It is only as we increase the strength of the driving field that the eigenphases undergo a repulsion which we refer to as an ``avoided crossing''. We describe 
quantitatively the ``avoided crossing'' between the eigenphases 
$\theta_{1}$ and $\theta_{2}$ in terms of occupation probabilities. In Fig.(13a) we plot 
the occupation probabilities $P_{0,1}$, $P_{1,1}$ and $P_{2,1}$ of the $|\theta_{1}>$
eigenvector on the propagating channels n=0,1,2 and in Fig.(13b) we plot
the occupation probabilities $P_{0,2}$, $P_{1,2}$ and $P_{2,2}$ of the $|\theta_{2}>$
eigenvector on the propagating channels n=0,1,2 as a function of Floquet energy
${\mathcal{E}}$. Before the ``avoided crossing'', ${\mathcal{E}}<0.106$, $|\theta_{1}>$ has support mainly on the second propagating channel,
 $P_{1,1}\approx 0.5$, and $|\theta_{2}>$ on the first propagating channel, $P_{0,2}\approx 
0.5$, while after the ``avoided crossing'', ${\mathcal{E}}>0.106$, $|\theta_{1}>$ has support mainly on 
the first propagating channel,
 $P_{0,1}\approx 0.5$, and $|\theta_{2}>$ on the second propagating channel, $P_{1,2}\approx 
0.5$. This total exchange of character is what we refer to as a {\it{sharp}} ``avoided crossing''.  
 Note that for $\alpha_{0}=0.5$ the propagating channels involved in the ``avoided crossing''
are mainly $n=0,1$. As the strength of the driving field $\alpha_{0}$ is increased an increasing
number of propagating channels undergo ``avoided crossings'', as shown in Tables II and III.
       
 For increased strength of the driving field the number of avoided crossings increases, see 
Figs.(\ref{fig:phases125}), (\ref{fig:phases225}) and it can be that more than two 
eigenphases participate in an ``avoided crossing'' for a certain Floquet energy. For 
example, this is the case for the ``avoided crossing'' at ${\mathcal{E}}\approx 0.14$, for 
$\alpha_{0}=2.25$, where there are three eigenphases $\theta_{1}$, $\theta_{2}$ and 
$\theta_{6}$ interfering, see Fig.(\ref{fig:phases225}). In Figs.(\ref{fig:phases125})
and (\ref{fig:phases225}) we plot the eigenphases of the Floquet 
$\mbox{\boldmath$S$}$-matrix as a function of ${\mathcal{E}}$ for $\alpha_{0}=1.25$ and 
$\alpha_{0}=2.25$, respectively, to show the increase in the number of ``avoided crossings''
with increasing strength of the driving field. In Tables II and III we present for 
$\alpha_{0}=1.25$ and $\alpha_{0}=2.25$, respectively, the eigephases which
undergo ``avoided crossings'' for different Floquet energies ${\mathcal{E}}$ and 
the propagating 
channels $n=0,1,...$ for which the occupation probability $P_{n,i}$ is substantial.
 In Tables II and III, the channels with small occupation probabilities are indicated as subscripts
to the channels with large occupation probabilities.  
 This is only an approximate picture but helps us visualize how the eigenphases change 
character at the ``avoided crossings''. For example, from table III, we obtain an approximate
picture how the eigenphases  $\theta_{1}$, $\theta_{2}$ and $\theta_{6}$ participate
in the ``avoided crossing'' at ${\mathcal{E}}\approx 0.14$. For ${\mathcal{E}}=0.07$, the eigenvector $|\theta_{1}>$
has support mainly on the propagating channel $n=1$ and less on $n=0,2,3$, the eigenvector 
$|\theta_{2}>$ has support mainly on the propagating channel $n=0$ and less on $n=1$, the
eigenvector $|\theta_{6}>$
has support mainly on the propagating channel $n=3$ and less on $n=5,1$. For 
${\mathcal{E}}=0.145$, the 
eigenvector $|\theta_{1}>$
has support mainly on the propagating channel $n=0$ and less on $n=1$, the eigenvector 
$|\theta_{2}>$ has support mainly on the propagating channels $n=1,3$ and less on $n=5$, the
eigenvector $|\theta_{6}>$
has support mainly on the propagating channels $n=1,3$. Thus, there is 
an exchange of character among the eigenphases $\theta_{1}$, $\theta_{2}$ and
$\theta_{6}$ expressed in terms of the mainly interfering channels $n=0,1,3,5$ but it is not a complete exchange as in the case of the {\it{sharp}}
``avoided crossing'' at $\alpha_{0}=0.5$, see Fig.(\ref{fig:phases05}).
 The ``avoided crossings'' we have just described 
for the open quantum system under consideration, the inverted Gaussian
in the presence of a driving field,    
 are analogous
to what was seen in a bounded chaotic system \cite{bound} where the authors also discuss
two different types of ``avoided crossings''.      

 We now turn to the classical dynamics
of the inverted Gaussian potential in the presence of the driving field.
 Figs.(\ref{fig:px})
are strobe plots of the phase space dynamics, for constant frequency 
$\omega=0.236$ and increasing strength of the driving field, $\alpha_{0}$ is 
equal to $0.5$, $1.25$ and $2.25$. The strobe plots
are drawn by evolving a set of trajectories, with different initial conditions,
 and plotting the location of each trajectory at time intervals
 ($t_{n}=2\pi m/\omega$, $m=1,2,...$) equal to the period of the driving field. We 
indicate the location of the period-1 periodic orbits with filled squares. The
strobe plots are drawn in the Lab frame, see Eqs.(\ref{eq:Schrod}), 
 (\ref{eq:pot}) and (\ref{eq:A(t)}), and are exactly the same with those in
the K-H frame except that in the Lab frame the $x$ axis is shifted by 
$\alpha_{0}$ \cite{Henseler}.
 If no driving field is
present, the motion is regular and bounded for negative energies, while
it is unbounded for positive energies. When the driving field is turned on,
 the KAM tori in the regular island around $x=0$, $p=0$ 
start breaking up as $\alpha_{0}$ is increased and chaotic motion sets in. 
For $\alpha_{0}=0.5$, see Fig.(16a), the classical phase space
is mixed. There are two islands around the two stable periodic orbits
but there are also chaotic trajectories. As $\alpha_{0}$ is further increased
the remaining islands are very small, see Fig.(16b), until they totally 
disappear, see Fig.(16c), and the phase space in the scattering region becomes 
dominated by chaos.
 In addition, in Figs.(17a) and (17b)
where the initial values of the classical momenta are chosen to correspond to the middle of 
the Floquet propagating channels, we find that as $\alpha_{0}$ is increased more trajectories 
get pulled into the chaotic region of the classical phase space.
 Correspondingly, in the quantum treatment of the scattering problem we have seen
that as the strength of the driving field is increased the eigenphases of the 
Floquet $\mbox{\boldmath$S$}$-matrix undergo an increasing number of ``avoided crossings'' where 
more Floquet channels contribute to the scattering process.
 We thus believe that the ``avoided crossings'' 
are a quantum manifestation of the 
breaking of the constants of motion and chaos setting in in the classical 
phase space.

\section{Conclusions}

In this paper, we have studied the scattering of electron waves from an inverted
Gaussian potential, used to model the atomic potential, in the presence of strong
time periodic electric fields. Using Floquet theory, we have constructed the Floquet
$\mbox{\boldmath$S$}$-matrix in the K-H frame, where asymptotic states can be defined.
 We have computed the transmission resonances, for different strengths of the driving
field, and shown that they are associated with
zero-pole pairs of the Floquet $\mbox{\boldmath$S$}$-matrix in the complex energy plane.
 We have also computed the Wigner-Smith delay times which is a different way to ``probe'' the 
complex spectrum of the open quantum system.
 Finally, we have shown that the eigenphases of the open quantum system undergo
a number of ``avoided crossings'' as a function of the electron Floquet 
energy, 
 that increases with increasing strength of the driving field. We believe that the ``avoided crossings'' are quantum manifestations
of the destruction of the KAM surfaces and the onset of chaos in the classical phase
space.

\vspace{0.5cm}
{\bf{Acknowledgement}}: We wish to thank the Welch Foundation, Grant No F-1051,
 NSF Grant INT-9602971, and DOE contract No. DE-FG03-94ER14405 for partial 
support of this work. We also thank the University of Texas at Austin High 
Performance
Computing Center for use of their computer facilities. Finally the authors thank
A. Gursoy, T. Timberlake and A. Shaji for helpful discussions.

\begin{center}
{\bf{Appendix A}}
\end{center}

In what follows starting from Eqs.(\ref{eq:1}), (\ref{eq:2}), (\ref{eq:3})
and (\ref{eq:4}) we obtain the outgoing probability amplitudes of the 
propagating modes in terms of the incoming probability amplitudes of the 
propagating modes. 

Using $\mbox{\boldmath $K$}_{+}^{-1}=\mbox{\boldmath $K$}_{-}$ and
$\mbox{\boldmath $K$}_{+}'^{-1}\mbox{\boldmath $K$}_{-}'=
-\mbox{\boldmath $K$}_{-}^{2}$, we eliminate $\mbox{\boldmath $B$}^{out}$
from Eqs.(\ref{eq:1}) and (\ref{eq:2}) and obtain

\begin{equation}
\label{eq:5}
2\mbox{\boldmath $K$}_{-}^{2}\mbox{\boldmath $J$}\mbox{\boldmath $B$}^{in}=
\mbox{\boldmath $\mathcal{N}$}^{-1}\mbox{\boldmath $L$}_{1}\mbox{\boldmath $C$}
+\mbox{\boldmath $\mathcal{N}$}^{-1}\mbox{\boldmath $L$}_{2}\mbox{\boldmath $D$},
\end{equation}
where $\mbox{\boldmath$L$}_{1}=\mbox{\boldmath $K$}_{-}\mbox{\boldmath $X$}^{(1)}_{+}-
{\mbox{\boldmath $K$}_{+}'}^{-1}{\mbox{\boldmath $X$}^{(1)}}^{'}_{+}$ and
 $\mbox{\boldmath$L$}_{2}=\mbox{\boldmath $K$}_{-}\mbox{\boldmath $X$}^{(2)}_{+}-
{\mbox{\boldmath $K$}_{+}'}^{-1}{\mbox{\boldmath $X$}^{(2)}}^{'}_{+}$.
Using $\mbox{\boldmath $K$}_{+}^{-1}=\mbox{\boldmath $K$}_{-}$ and
${\mbox{\boldmath $K$}_{+}'}^{-1}\mbox{\boldmath $K$}'_{-}=
-\mbox{\boldmath $K$}_{-}^{2}$, we eliminate $\mbox{\boldmath $A$}^{out}$
from Eqs.(\ref{eq:3}) and (\ref{eq:4}) and obtain

\begin{equation}
\label{eq:6}
2\mbox{\boldmath $K$}_{-}^{2}\mbox{\boldmath $J$}\mbox{\boldmath $A$}^{in}=
\mbox{\boldmath $\mathcal{N}$}^{-1}\mbox{\boldmath $L$}_{3}\mbox{\boldmath $C$}
+\mbox{\boldmath $\mathcal{N}$}^{-1}\mbox{\boldmath $L$}_{4}\mbox{\boldmath $D$},
\end{equation}
where $\mbox{\boldmath$L$}_{3}=\mbox{\boldmath $K$}_{-}\mbox{\boldmath $X$}^{(1)}_{-}+
{\mbox{\boldmath $K$}_{+}'}^{-1}{\mbox{\boldmath $X$}^{(1)}}^{'}_{-}$ and
 $\mbox{\boldmath$L$}_{4}=\mbox{\boldmath $K$}_{-}\mbox{\boldmath $X$}^{(2)}_{-}+
{\mbox{\boldmath $K$}_{+}'}^{-1}{\mbox{\boldmath $X$}^{(2)}}^{'}_{-}$.
From Eqs.(\ref{eq:5}) and (\ref{eq:6}) we express $\mbox{\boldmath $C$}$,  
 $\mbox{\boldmath $D$}$ in terms of $\mbox{\boldmath $A$}^{in}$ and 
$\mbox{\boldmath $B$}^{in}$ as follows

\begin{equation}
\label{eq:7}
\mbox{\boldmath $C$}=2\mbox{\boldmath $G$}
[\mbox{\boldmath $L$}_{2}^{-1}\mbox{\boldmath $K$}_{-}^{2}\mbox{\boldmath $J$}\mbox{\boldmath $\mathcal{N}$}
\mbox{\boldmath $B$}^{in}-
\mbox{\boldmath $L$}_{4}^{-1}\mbox{\boldmath $K$}_{-}^{2}\mbox{\boldmath $J$}\mbox{\boldmath $\mathcal{N}$}
\mbox{\boldmath $A$}^{in}],
\end{equation}

\begin{equation}
\label{eq:8}
\mbox{\boldmath $D$}=2\mbox{\boldmath $H$}
[\mbox{\boldmath $L$}_{1}^{-1}\mbox{\boldmath $K$}_{-}^{2}\mbox{\boldmath $J$}\mbox{\boldmath $\mathcal{N}$}
\mbox{\boldmath $B$}^{in}-
\mbox{\boldmath $L$}_{3}^{-1}\mbox{\boldmath $K$}_{-}^{2}\mbox{\boldmath $J$}\mbox{\boldmath $\mathcal{N}$}
\mbox{\boldmath $A$}^{in}],
\end{equation}
where $\mbox{\boldmath$G$}=[\mbox{\boldmath $L$}_{2}^{-1}\mbox{\boldmath $L$}_{1}-
\mbox{\boldmath $L$}_{4}^{-1}\mbox{\boldmath $L$}_{3}]^{-1}$
and $\mbox{\boldmath$H$}=[\mbox{\boldmath $L$}_{1}^{-1}\mbox{\boldmath $L$}_{2}-
\mbox{\boldmath $L$}_{3}^{-1}\mbox{\boldmath $L$}_{4}]^{-1}$.
 Substituting Eqs.(\ref{eq:7}) and (\ref{eq:8}) in Eqs.(\ref{eq:1}) and 
(\ref{eq:3}) yields $\mbox{\boldmath $A$}^{out}$ and $\mbox{\boldmath $B$}^{out}$ in terms of $\mbox{\boldmath $A$}^{in}$ and $\mbox{\boldmath $B$}^{in}$

\begin{eqnarray}
\label{eq:9}
\mbox{\boldmath $A$}^{out}&=&\mbox{\boldmath $\mathcal{N}$}^{-1}
\mbox{\boldmath $r$}'\mbox{\boldmath $\mathcal{N}$}\mbox{\boldmath $A$}^{in}+
\mbox{\boldmath $\mathcal{N}$}^{-1}
\mbox{\boldmath $t$}\mbox{\boldmath $\mathcal{N}$}\mbox{\boldmath $B$}^{in}\nonumber\\
\mbox{\boldmath $B$}^{out}&=&\mbox{\boldmath $\mathcal{N}$}^{-1}
\mbox{\boldmath $t$}'\mbox{\boldmath $\mathcal{N}$}\mbox{\boldmath $A$}^{in}+
\mbox{\boldmath $\mathcal{N}$}^{-1}
\mbox{\boldmath $r$}\mbox{\boldmath $\mathcal{N}$}\mbox{\boldmath $B$}^{in},
\end{eqnarray}
or equivalently

\begin{eqnarray}
\label{eq:9b}
\mbox{\boldmath $\mathcal{N}$}\mbox{\boldmath $A$}^{out}&=&
\mbox{\boldmath $r$}'\mbox{\boldmath $\mathcal{N}$}\mbox{\boldmath $A$}^{in}+
\mbox{\boldmath $t$}\mbox{\boldmath $\mathcal{N}$}\mbox{\boldmath $B$}^{in}\nonumber\\
\mbox{\boldmath $\mathcal{N}$}\mbox{\boldmath $B$}^{out}&=&
\mbox{\boldmath $t$}'\mbox{\boldmath $\mathcal{N}$}\mbox{\boldmath $A$}^{in}+
\mbox{\boldmath $r$}\mbox{\boldmath $\mathcal{N}$}\mbox{\boldmath $B$}^{in},
\end{eqnarray}
where
\begin{eqnarray}
\label{eq:10}
\mbox{\boldmath $r$}'&=&-[\mbox{\boldmath $K$}_{-}^{2}+2\mbox{\boldmath $K$}_{-}
\mbox{\boldmath $X$}_{-}^{(1)}\mbox{\boldmath$G$}\mbox{\boldmath $L$}_{4}^{-1}\mbox{\boldmath $K$}_{-}^{2}\nonumber
+2\mbox{\boldmath $K$}_{-}
\mbox{\boldmath $X$}_{-}^{(2)}\mbox{\boldmath $H$}\mbox{\boldmath $L$}_{3}^{-1}\mbox{\boldmath $K$}_{-}^{2}]\mbox{\boldmath $J$}\nonumber\\
\mbox{\boldmath $r$}&=&[-\mbox{\boldmath $K$}_{-}^{2}+2\mbox{\boldmath $K$}_{-}
\mbox{\boldmath $X$}_{+}^{(1)}\mbox{\boldmath$G$}\mbox{\boldmath $L$}_{2}^{-1}\mbox{\boldmath $K$}_{-}^{2}
+2\mbox{\boldmath $K$}_{-}
\mbox{\boldmath $X$}_{+}^{(2)}\mbox{\boldmath $H$}\mbox{\boldmath $L$}_{1}^{-1}\mbox{\boldmath $K$}_{-}^{2}]\mbox{\boldmath $J$}\nonumber\\
\mbox{\boldmath $t$}'&=&-2[\mbox{\boldmath $K$}_{-}
\mbox{\boldmath $X$}_{+}^{(1)}\mbox{\boldmath $G$}\mbox{\boldmath $L$}_{4}^{-1}\mbox{\boldmath $K$}_{-}^{2}
+\mbox{\boldmath $K$}_{-}
\mbox{\boldmath $X$}_{+}^{(2)}\mbox{\boldmath $H$}\mbox{\boldmath $L$}_{3}^{-1}\mbox{\boldmath $K$}_{-}^{2}]\mbox{\boldmath $J$}\nonumber\\
\mbox{\boldmath $t$}&=&2[\mbox{\boldmath $K$}_{-}
\mbox{\boldmath $X$}_{-}^{(1)}\mbox{\boldmath $G$}\mbox{\boldmath $L$}_{2}^{-1}\mbox{\boldmath $K$}_{-}^{2}
+\mbox{\boldmath $K$}_{-}
\mbox{\boldmath $X$}_{-}^{(2)}\mbox{\boldmath $H$}\mbox{\boldmath $L$}_{1}^{-1}\mbox{\boldmath $K$}_{-}^{2}]\mbox{\boldmath $J$}
\end{eqnarray}
From Eqs.(\ref{eq:10}), due to the multiplication on the right by the $N\times N$
matrix $\mbox{\boldmath $J$}$, we find that the $N\times N$ matrices
$\mbox{\boldmath $r$}'$, $\mbox{\boldmath $r$}$, 
$\mbox{\boldmath $t'$}$ and $\mbox{\boldmath $t$}$ are of the following
form

\begin{equation}
\label{eq:11}
    \begin{array}{cccc}
     \mbox{\boldmath $r$}'=\left(\begin{array}{cc}

                      \mbox{\boldmath $0$}_{ee} & \mbox{\boldmath $r$}'_{ep}\\ 
                      \mbox{\boldmath $0$}_{pe} & \mbox{\boldmath $r$}'_{pp}
\end{array}\right), &
\mbox{\boldmath $r$}=\left(\begin{array}{cc}

                      \mbox{\boldmath $0$}_{ee} & \mbox{\boldmath $r$}_{ep}\\ 
                      \mbox{\boldmath $0$}_{pe} & \mbox{\boldmath $r$}_{pp}
\end{array}\right),&
\mbox{\boldmath $t$}'=\left(\begin{array}{cc}

                      \mbox{\boldmath $0$}_{ee} & \mbox{\boldmath $t$}'_{ep}\\ 
                      \mbox{\boldmath $0$}_{pe} & \mbox{\boldmath $t$}'_{pp}
\end{array}\right),&
\mbox{\boldmath $t$}=\left(\begin{array}{cc}

                      \mbox{\boldmath $0$}_{ee} & \mbox{\boldmath $t$}_{ep}\\ 
                      \mbox{\boldmath $0$}_{pe} & \mbox{\boldmath $t$}_{pp}
\end{array}\right),\end{array}
\end{equation} 
where the matrices $\mbox{\boldmath $r$}'_{ep}$, $\mbox{\boldmath $r$}_{ep}$,
 $\mbox{\boldmath $t$}'_{ep}$ and $\mbox{\boldmath $t$}_{ep}$
have dimensions $n_{e}\times n_{p}+1$ and the matrices 
$\mbox{\boldmath $r$}'_{pp}$, $\mbox{\boldmath $r$}_{pp}$,
 $\mbox{\boldmath $t$}'_{pp}$ and $\mbox{\boldmath $t$}_{pp}$
have dimensions 
$n_{p}+1\times n_{p}+1$, respectively, ($n_{e}$ is the number of the evanescent modes and $n_{p}+1$ is the number of the propagating modes). 
 The matrices $\mbox{\boldmath$0$}_{ee}$ and $\mbox{\boldmath$0$}_{pe}$
have dimensions $n_{e}\times n_{e}$ and $n_{p}+1\times n_{e}$, respectively,
and they have zero elements because the amplitudes 
$b^{in}$ and $a^{in}$ of the evanescent modes are zero,
 $b_{n}^{in}=a_{n}^{in}=0$, for $n=-n_{e},...,-1$.

In addition, the matrices $\mbox{\boldmath $\mathcal{N}$}$,
 $\mbox{\boldmath $A$}^{in}$, $\mbox{\boldmath $A$}^{out}$, 
 $\mbox{\boldmath $B$}^{in}$ and $\mbox{\boldmath $B$}^{out}$ can be written as
\begin{equation}
\label{eq:11b}
 \begin{array}{ccccc}   \mbox{\boldmath $\mathcal{N}$}=\left(\begin{array}{cc}

                      \mbox{\boldmath $\mathcal{N}$}_{ee} & \mbox{\boldmath $0$}_{ep}\\ 
                      \mbox{\boldmath $0$}_{pe} & \mbox{\boldmath $\mathcal{N}$}_{pp}
\end{array}\right),&
\mbox{\boldmath $A$}^{in}=\left(\begin{array}{c}

                      \mbox{\boldmath $A$}^{in}_{e} \\
                      \mbox{\boldmath $A$}^{in}_{p}
\end{array}\right),&
\mbox{\boldmath $A$}^{out}=\left(\begin{array}{c}

                      \mbox{\boldmath $A$}^{out}_{e} \\
                      \mbox{\boldmath $A$}^{out}_{p}
\end{array}\right),&
\mbox{\boldmath $B$}^{in}=\left(\begin{array}{c}

                      \mbox{\boldmath $B$}^{in}_{e} \\
                      \mbox{\boldmath $B$}^{in}_{p}
\end{array}\right),&
\mbox{\boldmath $B$}^{out}=\left(\begin{array}{c}

                      \mbox{\boldmath $B$}^{out}_{e} \\
                      \mbox{\boldmath $B$}^{out}_{p}
\end{array}\right)
 \end{array},
\end{equation} 
where the matrices $\mbox{\boldmath $\mathcal{N}$}_{ee}$, $\mbox{\boldmath $\mathcal{N}$}_{pp}$
have dimensions $n_{e}\times n_{e}$ and $n_{p}+1\times n_{p}+1$, respectively. 
 The elements of 
the $n_{e}\times 1$ matrices $\mbox{\boldmath $A$}^{in}_{e}$,
 $\mbox{\boldmath $A$}^{out}_{e}$, $\mbox{\boldmath $B$}^{in}_{e}$
and $\mbox{\boldmath $B$}^{out}_{e}$ are the amplitudes of the evanescent modes. 
 The elements of the $n_{p}+1\times 1$ matrices $\mbox{\boldmath $A$}^{in}_{p}$,
 $\mbox{\boldmath $A$}^{out}_{p}$, $\mbox{\boldmath $B$}^{in}_{p}$
and $\mbox{\boldmath $B$}^{out}_{p}$ are the amplitudes of the propagating modes.
 Using Eqs.(\ref{eq:11}) and (\ref{eq:11b}) we write Eqs.(\ref{eq:9b}) as follows
\begin{eqnarray}
\label{eq:12}
\mbox{\boldmath $\mathcal{N}$}_{ee}\mbox{\boldmath $A$}_{e}^{out}&=&
\mbox{\boldmath $r$}_{ep}'\mbox{\boldmath $\mathcal{N}$}_{pp}\mbox{\boldmath $A$}_{p}^{in}+\mbox{\boldmath $t$}_{ep}\mbox{\boldmath $\mathcal{N}$}_{pp}\mbox{\boldmath $B$}_{p}^{in}\nonumber\\
\mbox{\boldmath $\mathcal{N}$}_{pp}\mbox{\boldmath $A$}_{p}^{out}&=&
\mbox{\boldmath $r$}_{pp}'\mbox{\boldmath $\mathcal{N}$}_{pp}\mbox{\boldmath $A$}_{p}^{in}+\mbox{\boldmath $t$}_{pp}\mbox{\boldmath $\mathcal{N}$}_{pp}\mbox{\boldmath $B$}_{p}^{in}\nonumber\\
\mbox{\boldmath $\mathcal{N}$}_{ee}\mbox{\boldmath $B$}_{e}^{out}&=&
\mbox{\boldmath $t$}_{ep}'\mbox{\boldmath $\mathcal{N}$}_{pp}\mbox{\boldmath $A$}_{p}^{in}+\mbox{\boldmath $r$}_{ep}\mbox{\boldmath $\mathcal{N}$}_{pp}\mbox{\boldmath $B$}_{p}^{in}\nonumber\\
\mbox{\boldmath $\mathcal{N}$}_{pp}\mbox{\boldmath $B$}_{p}^{out}&=&
\mbox{\boldmath $t$}_{pp}'\mbox{\boldmath $\mathcal{N}$}_{pp}\mbox{\boldmath 
$A$}_{p}^{in}+\mbox{\boldmath $r$}_{pp}\mbox{\boldmath $\mathcal{N}$}_{pp}\mbox{\boldmath $B$}_{p}^{in}.
\end{eqnarray}
From Eqs.(\ref{eq:12}) we obtain the Floquet $\mbox{\boldmath$S$}$-matrix
given in Eq.(\ref{eq:13}).

\begin{center}
{\bf{Appendix B}}
\end{center}

In sections II.B, C and D we have formally constructed the Floquet 
$\mbox{\boldmath$ S$}$-matrix in terms of the functions $\chi_{n,j}(x)$, 
with $n=-n_{e},...,0,...,n_{p}$ and $j=1,...,2N$, which are linearly independent
functions in the scattering region II. For the inverted Gaussian potential
the functions $\chi_{n,j}(x)$
can only be obtained numerically. In Appendix A we formally expressed 
the matrices $\mbox{\boldmath$r$}$ and $\mbox{\boldmath$t$}$ in terms of the 
functions $\chi_{n,j}(x)$.
Numerically, though, it is not efficient to compute the functions $\chi_{n,j}(x)$.  
 In what follows we outline the numerical
method \cite{Bhatt} we use to obtain the $N\times N$ matrices 
$\mbox{\boldmath $r$}$ and
$\mbox{\boldmath $t$}$ for electron waves incident from the right.
 
The wavefunction in the asymptotic regions I and III is given by 
Eqs.(\ref{eq:solutI}) and (\ref{eq:solutIII}) with $a_{n}^{in}=0$, since we 
only consider electron waves incident from the right. We can then write

\begin{equation}
\label{eq:App1}
\begin{array}{c}
\phi_{n}(x_{0})=b_{n}^{out}\frac{e^{i k_{n} x_{0}}}{\sqrt{k_{n}}}+b_{n}^{in}
\frac{e^{-i k_{n}x_{0}}}{\sqrt{k_{n}}},\\
\phi_{n}(-x_{0})=a_{n}^{out}\frac{e^{i k_{n} x_{0}}}{\sqrt{k_{n}}}.\end{array}
\end{equation}
with $b_{n}^{in}=0$ for $n=-n_{e},...,-1$. 
 Next, we write Eqs.(\ref{eq:App1}) in matrix form using
Eqs.(\ref{eq:matrix1}), (\ref{eq:matrix3}) and (\ref{eq:matrix3a}) as follows

\begin{equation}
\label{eq:App2}
\mbox{\boldmath $\phi$}(x_{0})=\mbox{\boldmath$K$}_{+}
\mbox{\boldmath $\mathcal{N}$}\mbox{\boldmath $B$}^{out}
+\mbox{\boldmath$K$}_{-}\mbox{\boldmath$J$}\mbox{\boldmath $\mathcal{N}$}
\mbox{\boldmath $B$}^{in},
\end{equation}

\begin{equation}
\label{eq:App3}
\mbox{\boldmath $\phi$}(-x_{0})=\mbox{\boldmath$K$}(-x_{0})
\mbox{\boldmath $\mathcal{N}$}\mbox{\boldmath $A$}^{out},
\end{equation}
where
\begin{equation}
\label{eq:matrix5}
\mbox{\boldmath $K$}(x)=\left(\begin{array}{ccccc}
                e^{- i k_{-n_{e}}x} &  &  & &    \\
                                       &\vdots& & &0  \\
                                        & &e^{- i k_{0}x}& &  \\
                                         & & &\vdots & \\
                                0   & & & &e^{- i k_{n_{p}}x}\end{array}\right).
\end{equation}
Using Eqs.(\ref{eq:9b}) with $\mbox{\boldmath$A$}^{in}=0$, we write
Eqs.(\ref{eq:App2}), (\ref{eq:App3}) as follows

 \begin{equation}
\label{eq:App4}
\mbox{\boldmath $\phi$}(x_{0})=\mbox{\boldmath$K$}_{+}
\mbox{\boldmath $r$}\mbox{\boldmath $\mathcal{N}$}\mbox{\boldmath $B$}^{in}
+\mbox{\boldmath$K$}_{-}
\mbox{\boldmath$J$}\mbox{\boldmath $\mathcal{N}$}\mbox{\boldmath $B$}^{in},
\end{equation}
\begin{equation}
\label{eq:App5}
\mbox{\boldmath $\phi$}(-x_{0})=\mbox{\boldmath$K$}(-x_{0})
\mbox{\boldmath $t$}\mbox{\boldmath $\mathcal{N}$}\mbox{\boldmath $B$}^{in}.
\end{equation}
 Next, we use Eq.(\ref{eq:trunsys}) to numerically propagate 
$\mbox{\boldmath $\phi$}(x)$ from $x=-x_{0}$ up to $x=x_{0}$ according to the
Numerov algorithm \cite{Numerov}.
 From 
Eq.(\ref{eq:App5}) we see that in practice we numerically integrate the
$\mbox{\boldmath$K$}(x)$ matrix from $x=-x_{0}$ up to
$x=x_{0}$, since $\mbox{\boldmath $t$}$, $\mbox{\boldmath $\mathcal{N}$}$ and $\mbox{\boldmath $B$}^{in}$ are constant matrices.
 Let us indicate by $\mbox{\boldmath$\tilde{K}$}(x_{0})$ the numerically integrated
matrix $\mbox{\boldmath$K$}(x)$ at $x=x_{0}$. Then, matching the wavefunction
and its first derivative at $x=x_{0}$ and using Eqs.(\ref{eq:matrix1}) we 
obtain
\begin{equation}
\label{eq:App7}
\begin{array}{c}
\mbox{\boldmath$\tilde{K}$}(x_{0})\mbox{\boldmath $t$}\mbox{\boldmath $\mathcal{N}$}\mbox{\boldmath $B$}^{in}=\mbox{\boldmath$K$}_{+}
\mbox{\boldmath $r$}\mbox{\boldmath $\mathcal{N}$}\mbox{\boldmath $B$}^{in}
+\mbox{\boldmath$K$}_{-}
\mbox{\boldmath$J$}\mbox{\boldmath $\mathcal{N}$}\mbox{\boldmath $B$}^{in}\\
\mbox{\boldmath$\tilde{K}$}'(x_{0})\mbox{\boldmath $t$}\mbox{\boldmath $\mathcal{N}$}\mbox{\boldmath $B$}^{in}=\mbox{\boldmath$K$}_{+}'
\mbox{\boldmath $r$}\mbox{\boldmath $\mathcal{N}$}\mbox{\boldmath $B$}^{in}
+\mbox{\boldmath$K$}_{-}'
\mbox{\boldmath$J$}\mbox{\boldmath $\mathcal{N}$}\mbox{\boldmath $B$}^{in},\end{array}
\end{equation}
or equivalently
\begin{equation}
\label{eq:App8}
\begin{array}{c}
\mbox{\boldmath$\tilde{K}$}(x_{0})\mbox{\boldmath $t$}=\mbox{\boldmath$K$}_{+}
\mbox{\boldmath $r$}
+\mbox{\boldmath$K$}_{-}
\mbox{\boldmath$J$}\\
\mbox{\boldmath$\tilde{K}$}'(x_{0})\mbox{\boldmath $t$}=\mbox{\boldmath$K$}_{+}'
\mbox{\boldmath $r$}
+\mbox{\boldmath$K$}_{-}'
\mbox{\boldmath$J$}.\end{array}
\end{equation}     
From Eqs.(\ref{eq:App8}) we find 
\begin{equation}
\label{eq:App9}
\begin{array}{c}
\mbox{\boldmath $r$}=[\mbox{\boldmath$\tilde{K}$}(x_{0})^{-1}\mbox{\boldmath$K$}_{+}-\mbox{\boldmath$\tilde{K}$}'(x_{0})^{-1}\mbox{\boldmath$K$}'_{+}]^{-1}
[\mbox{\boldmath$\tilde{K}$}(x_{0})^{-1}\mbox{\boldmath$K$}_{-}-\mbox{\boldmath$\tilde{K}$}'(x_{0})^{-1}\mbox{\boldmath$K$}'_{-}]\mbox{\boldmath$J$}\\
\mbox{\boldmath$t$}=[\mbox{\boldmath$K$}_{-}\mbox{\boldmath$\tilde{K}$}(x_{0})-
{\mbox{\boldmath$K$}_{+}'}^{-1}\mbox{\boldmath$\tilde{K}$}'(x_{0})]^{-1}
[\mbox{\boldmath$K$}_{-}^{2}-{\mbox{\boldmath$K$}_{+}'}^{-1}\mbox{\boldmath$K$}_{-}']\mbox{\boldmath$J$},\end{array}
\end{equation}
where we have used the relation $\mbox{\boldmath$K$}_{+}^{-1}=\mbox{\boldmath$K$}_{-}$.
The matrices $\mbox{\boldmath $r$}$ and $\mbox{\boldmath $t$}$ given by 
Eqs.(\ref{eq:App9}) are of the form shown in Eq.(\ref{eq:11}) and thus we
can extract the matrices $\mbox{\boldmath$r$}_{pp}$ and $\mbox{\boldmath$t$}_{pp}$.
 Following section II.D, we then obtain the matrices $\mbox{\boldmath $R$}$ and $\mbox{\boldmath $T$}$. 
 Finally, using the symmetry property of the Floquet 
$\mbox{\boldmath$S$}$-matrix given in Eq.(\ref{eq:symmetry}) we find the 
matrices $\mbox{\boldmath $R'$}$ and $\mbox{\boldmath $T'$}$ for electron 
waves incident from the left.

\newpage
\begin{center}
LIST OF TABLES
\end{center}
\begin{enumerate}
\item Table I: The Wigner-Smith delay times $\tau_{ws}^{n}$ compared to the lifetime
 $\tau_{L}$ for the 1st and 2nd quasibound states for $\alpha_{0}$
equal to $0.5$, $2.25$ and $5.25$.

\item Table II: For $\alpha_{0}=1.25$ we retain $n=-6,...,0,...,6$ channels and obtain
$14$ eigenphases from the $14\times 14$ Floquet $\mbox{\boldmath$S$}$-matrix.
 In Table II we only show the five eigenphases participating in the 
``avoided crossings'' at different Floquet energies ${\mathcal{E}}$, see 
Fig.(\ref{fig:phases125}).
 For each of the five participating eigenphases $\theta_{1},...,\theta_{5}$ 
we display the propagating 
channels $n=0,1,...$ with substantial occupation probability $P_{n,i}$. 
 The propagating channels 
involved in the ``avoided crossings'' are $n=0,1,2$.

\item Table III: For $\alpha_{0}=2.25$ we retain $n=-12,...,0,...,12$ channels and obtain
$25$ eigenphases from the $25\times 25$ Floquet $\mbox{\boldmath$S$}$-matrix.
 In Table III we only show the seven eigenphases participating in the 
``avoided crossings'' at different Floquet energies ${\mathcal{E}}$, see 
Fig.(\ref{fig:phases225}).
 For each of the seven participating eigenphases $\theta_{1},...,\theta_{7}$ 
we display the propagating 
channels $n=0,1,...$ with substantial occupation probability $P_{n,i}$. 
 The propagating channels 
involved in the ``avoided crossings'' are $n=0,1,...,5$.  
\end{enumerate}

\newpage
\begin{center}
Table I
\end{center}

\begin{center}
\begin{tabular}{ccc} \hline \hline
$Resonance$ & $\tau_{ws}^{n} (a.u.)$ &$\tau_{L} (a.u.)$ \\ \hline
$\alpha_{0}=0.5$     &            &   \\ 
\mbox{1st resonance}  &$390$       & $208$  \\ \hline 
$\alpha_{0}=2.25$     &            &     \\
\mbox{1st resonance}  &$43$       & $28.2$  \\
\mbox{2nd resonance}  &$137$       & $104$  \\  \hline
$\alpha_{0}=5.25$     &            &     \\
\mbox{1st resonance}  &$244\times 10$       & $704$  \\
\mbox{2nd resonance}  &$329$       & $81$  \\ \hline \hline
\end{tabular}
\end{center}

\begin{center}
Table II
\end{center}

\begin{center}
\begin{tabular}{|c|c|c|c|c|c|} \hline
${\mathcal{E}} (a.u.)$ & $\theta_{1}$ &$\theta_{2}$ &$\theta_{3}$ &$\theta_{4}$ &$\theta_{5}$\\ \hline
$0.01$      &  $0_{1}$   & $1_{0}$   & $0$       & $2$     &  $1$    \\ \hline
$0.07 $     &  $1_{0}$   & $0_{1}$   & $0$       & $2$     &  $1$    \\ \hline
$0.12 $     &  $0$       & $1$       & $0$       & $2$     &  $1$    \\ \hline
$0.15 $     &  $2$       & $1$       & $0$       & $0$     &  $1$    \\ \hline 
$0.23 $     &  $2$       & $1$       & $0_{1}$   & $0$     &  $1_{0}$ \\ \hline
\end{tabular}
\end{center}

\begin{center}
Table III
\end{center}

\begin{center}
\begin{tabular}{|c|c|c|c|c|c|c|c|} \hline
${\mathcal{E}} (a.u.)$ & $\theta_{1}$ &$\theta_{2}$ &$\theta_{3}$ &$\theta_{4}$ &$\theta_{5}$ & $\theta_{6}$ & $\theta_{7}$ \\ \hline
$0.01$    &$0_{1,2}$      &$1_{0,3}$   & $0_{1}$             &$2_{4}$       &$1_{0,2,3}$       &$3_{1,5}$ &$3_{1,5}$ \\ \hline
$0.07 $   &$1_{0_{2,3}}$   & $0_{1}$     & $0_{1_{2}}$   &$2_{4}$       & $1_{0,2,3}$  &$3_{5_{1}}$  &$3_{1,5}$ \\ \hline
$0.145 $  &$0_{1}$            & $1,3_{5}$   & $0_{1_{2}}$    &$2_{4}$       & $1_{0,2,3}$ &$1,3$ &$3_{1,5}$ \\ \hline
$0.18 $   &$2_{0,4}$       & $3_{1,5}$   & $0_{1_{2}}$   &$0_{2}$       & $1_{0,2,3}$ &$1_{3}$  &$3_{1,5}$\\ \hline 
$0.22 $   &$2_{0,4}$       & $3_{1,5}$   & $1_{0}$       &$0_{2_{1}}$       & $0_{1}$  &$1_{3}$ &$3_{1,5}$\\ \hline
$0.23 $   &$2_{0,4}$       & $3_{1,5}$   & $3_{1,5}$      & $0_{2_{1}}$       &$0_{2}$ &$1_{3}$  &$1_{3}$\\ \hline
\end{tabular}
\end{center}

\begin{figure}
\begin{centering}
\leavevmode
\epsfxsize=0.45\linewidth
\epsfbox{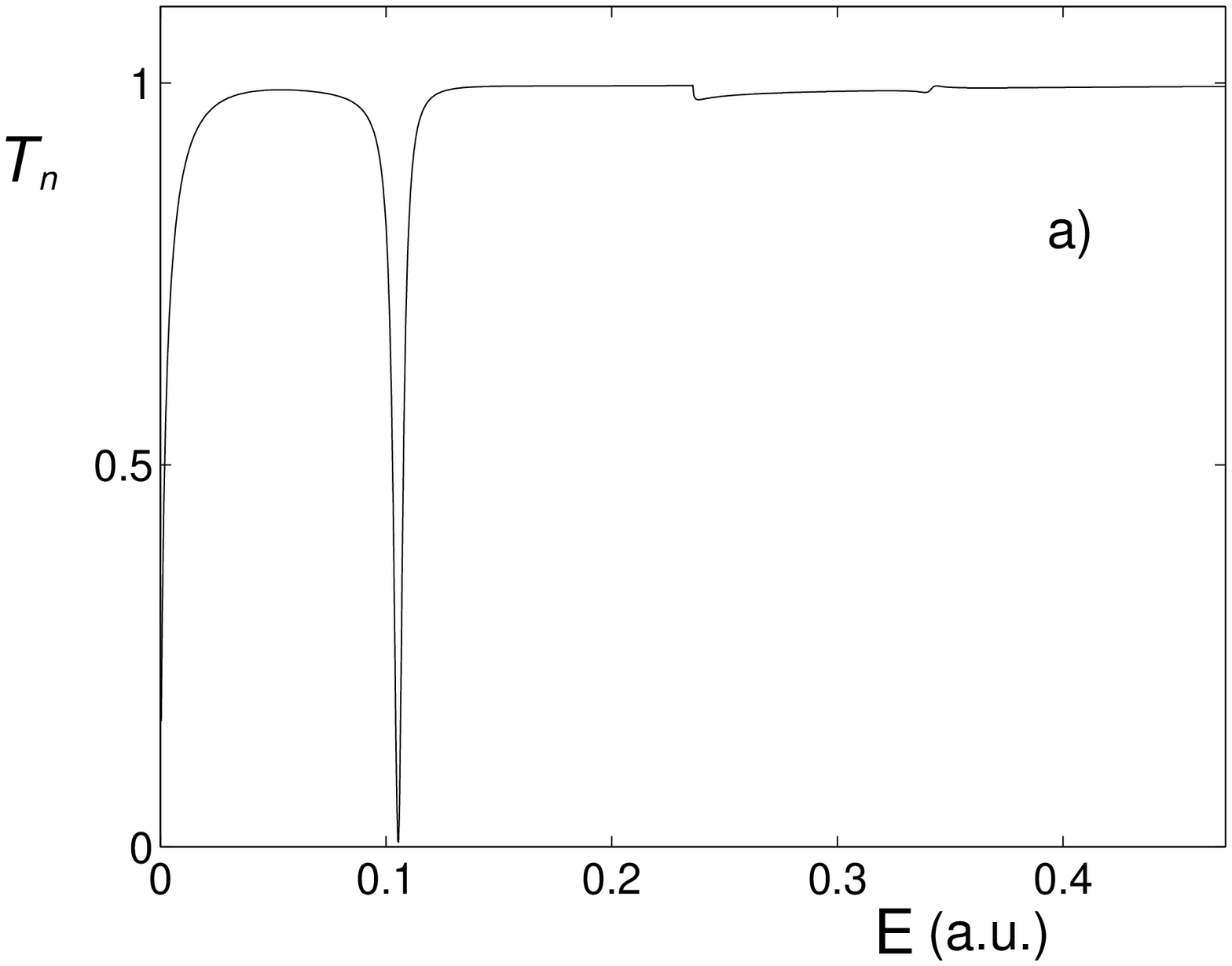}
\leavevmode
\epsfxsize=0.45\linewidth
\epsfbox{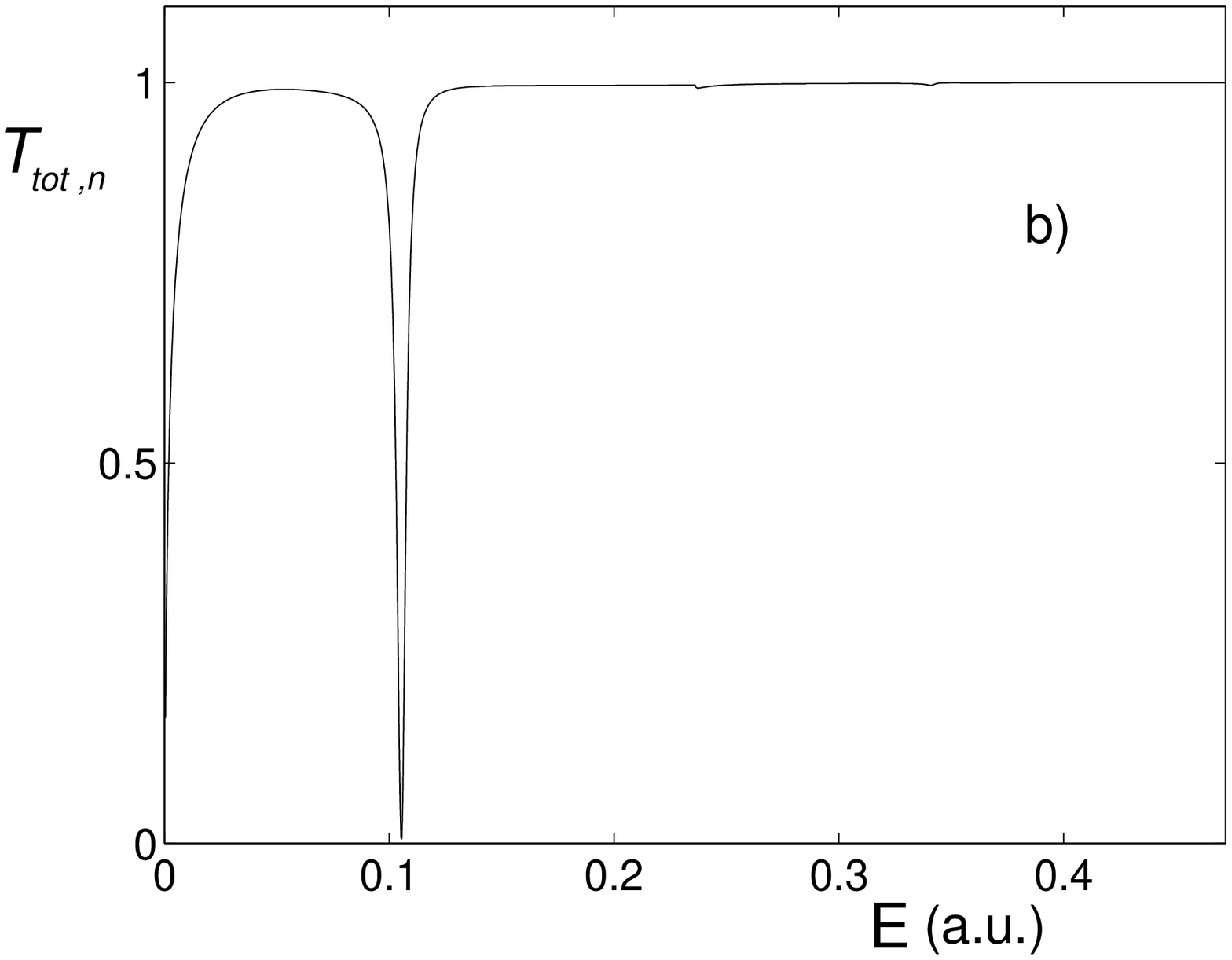}
\caption{}
\label{fig:T05}
\end{centering}
\end{figure}

\begin{figure}
\begin{centering}
\leavevmode
\epsfxsize=0.45\linewidth
\epsfbox{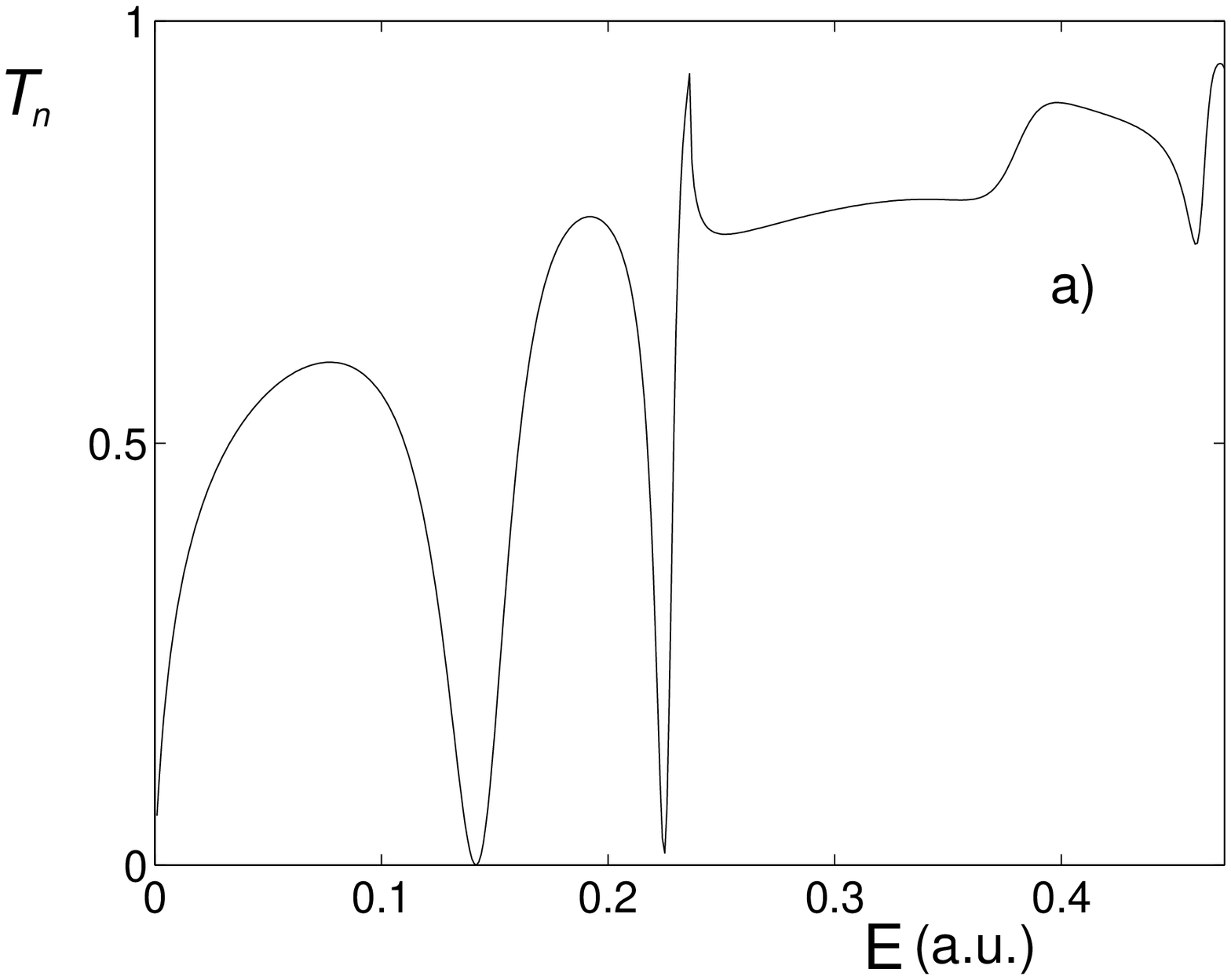}
\leavevmode
\epsfxsize=0.45\linewidth
\epsfbox{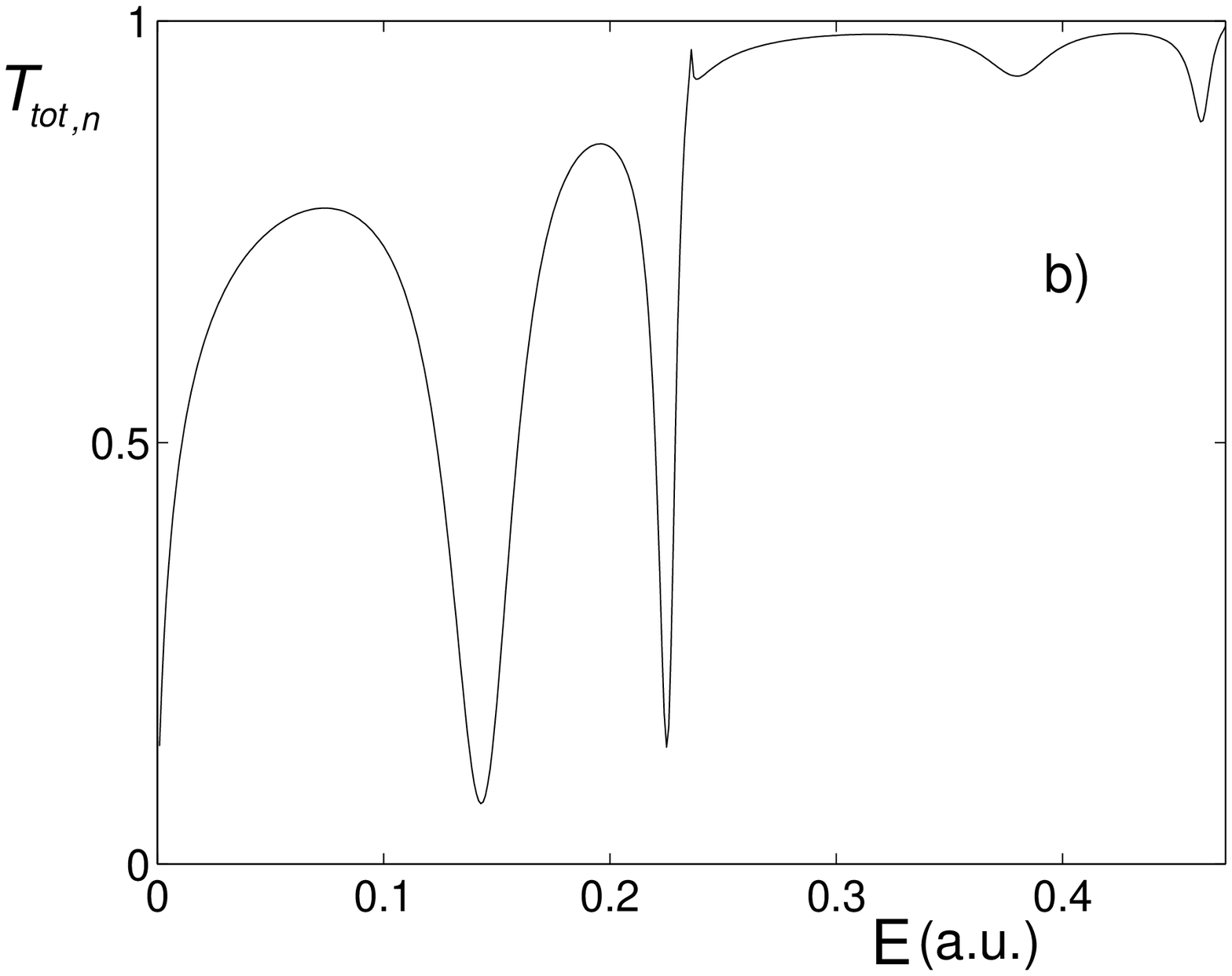}
\caption{}
\label{fig:T225}
\end{centering}
\end{figure}

\begin{figure}
\begin{centering}
\leavevmode
\epsfxsize=0.45\linewidth
\epsfbox{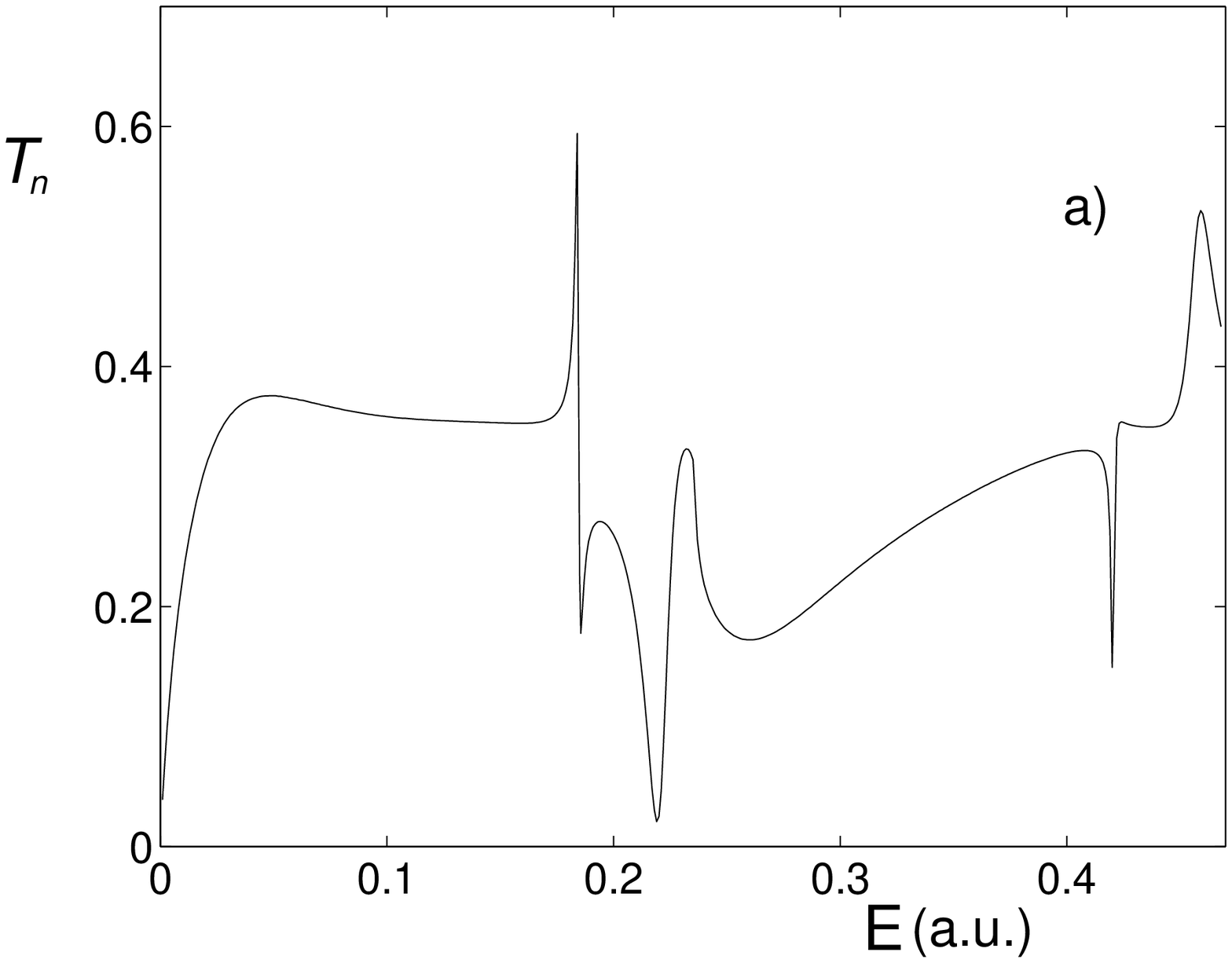}
\leavevmode
\epsfxsize=0.45\linewidth
\epsfbox{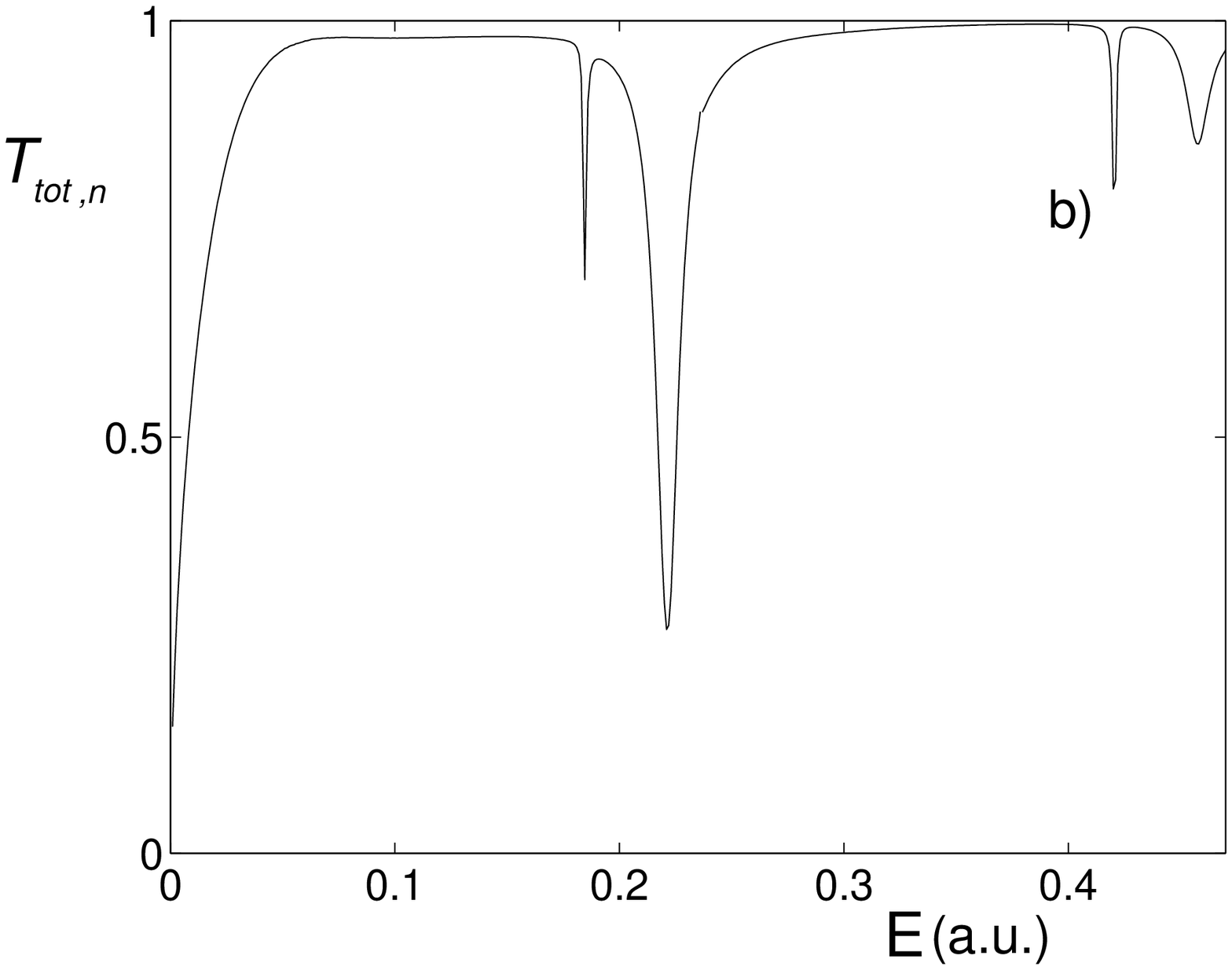}
\caption{}
\label{fig:T525}
\end{centering}
\end{figure}

\begin{figure}
\begin{centering}
\leavevmode
\epsfxsize=1\linewidth
\epsfbox{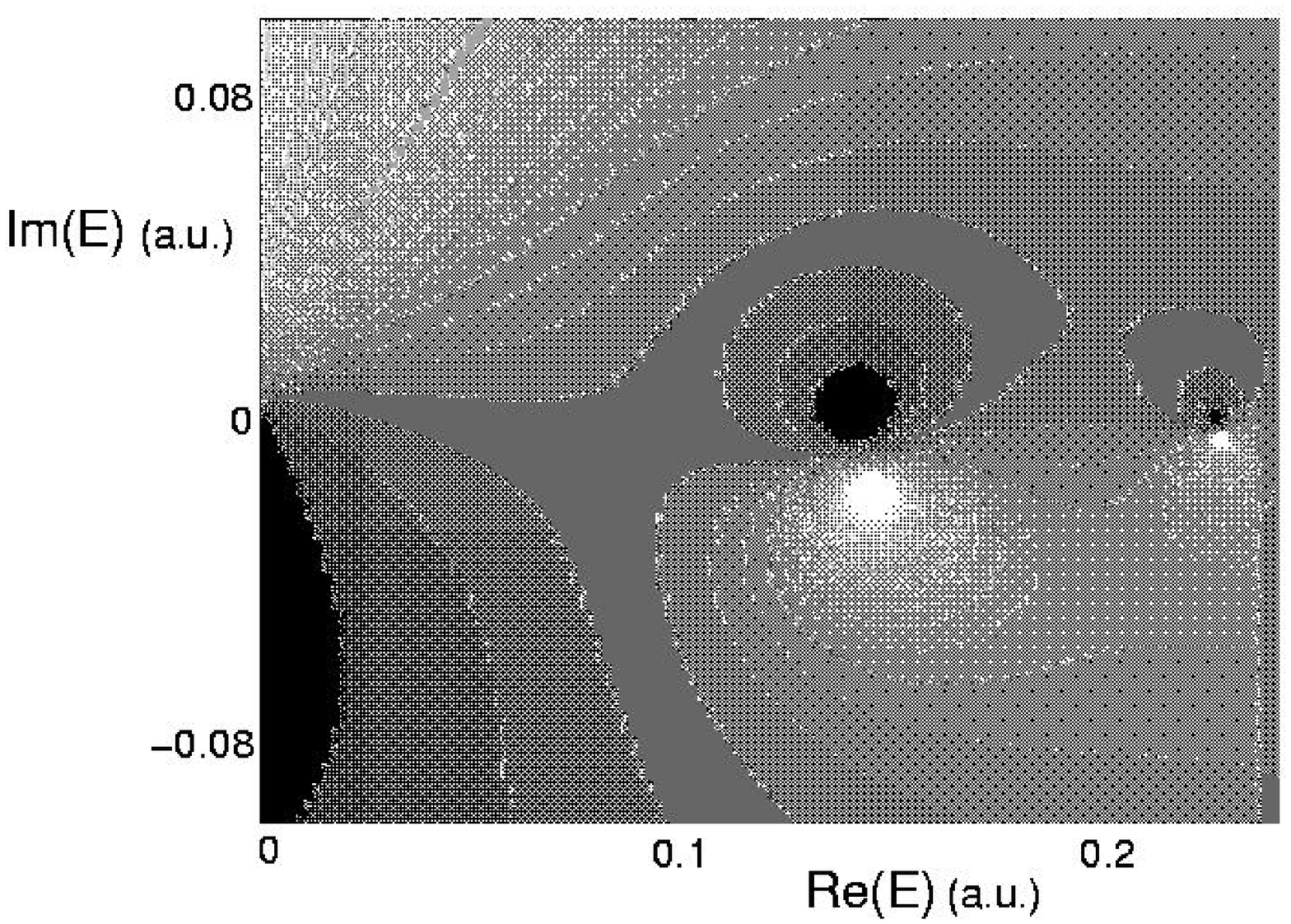}
\caption{}
\label{fig:225cont}
\end{centering}
\end{figure}

\begin{figure}
\begin{centering}
\leavevmode
\epsfxsize=0.5\linewidth
\epsfbox{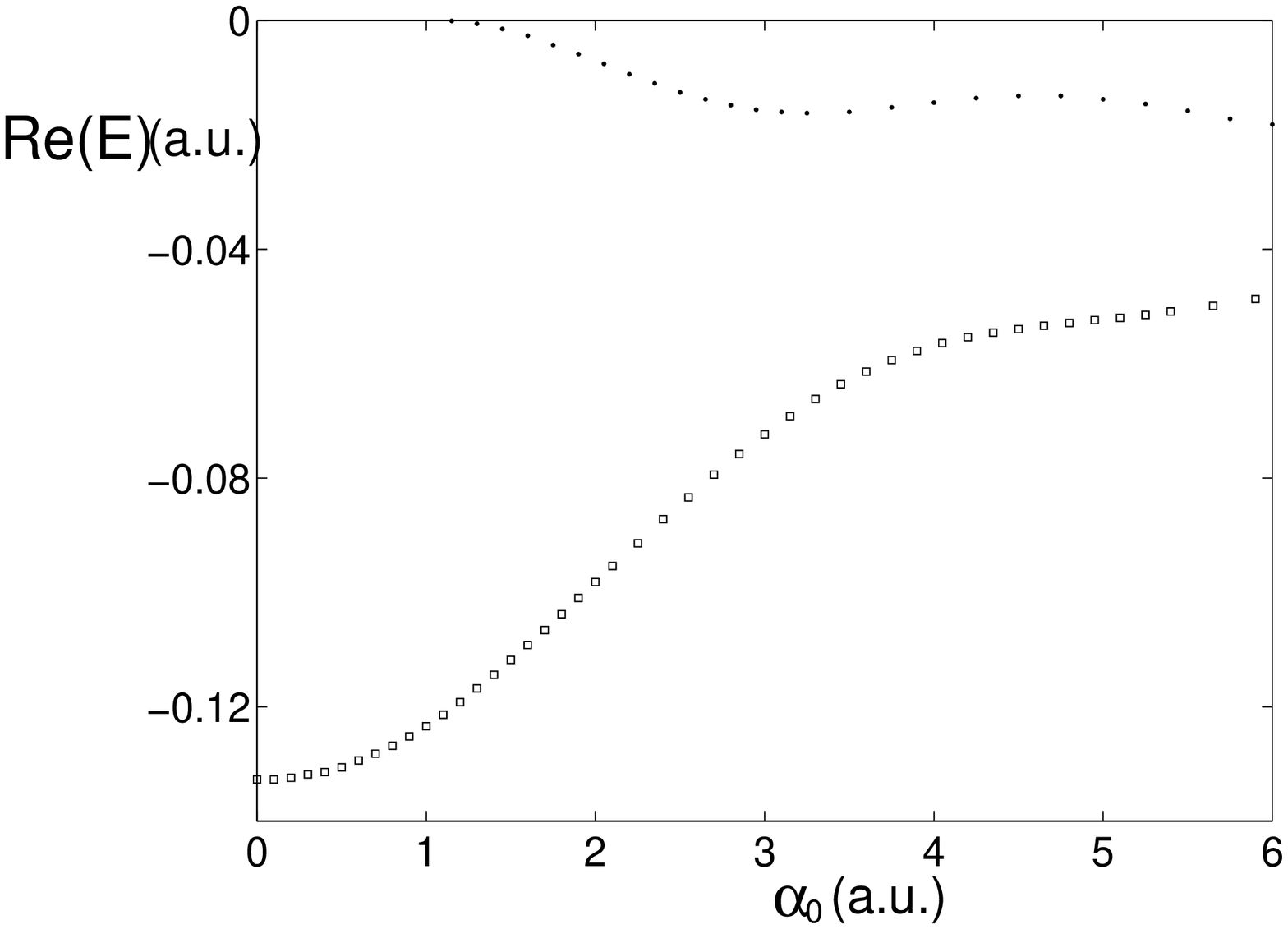}
\caption{}
\label{fig:energyr}
\end{centering}
\end{figure}

\begin{figure}
\begin{centering}
\leavevmode
\epsfxsize=0.5\linewidth
\epsfbox{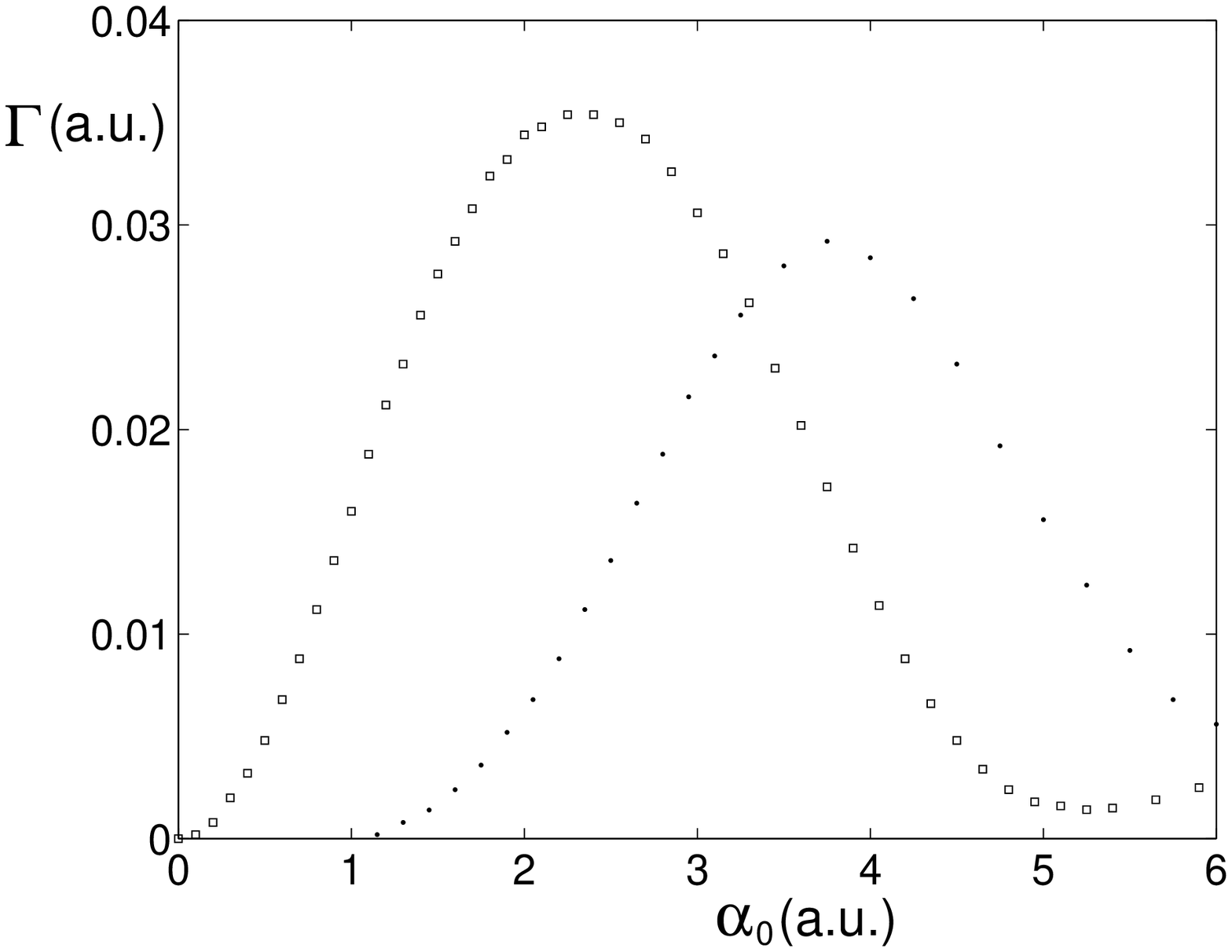}
\caption{}
\label{fig:energyi}
\end{centering}
\end{figure}

\begin{figure}
\begin{centering}
\leavevmode
\epsfxsize=0.45\linewidth
\epsfbox{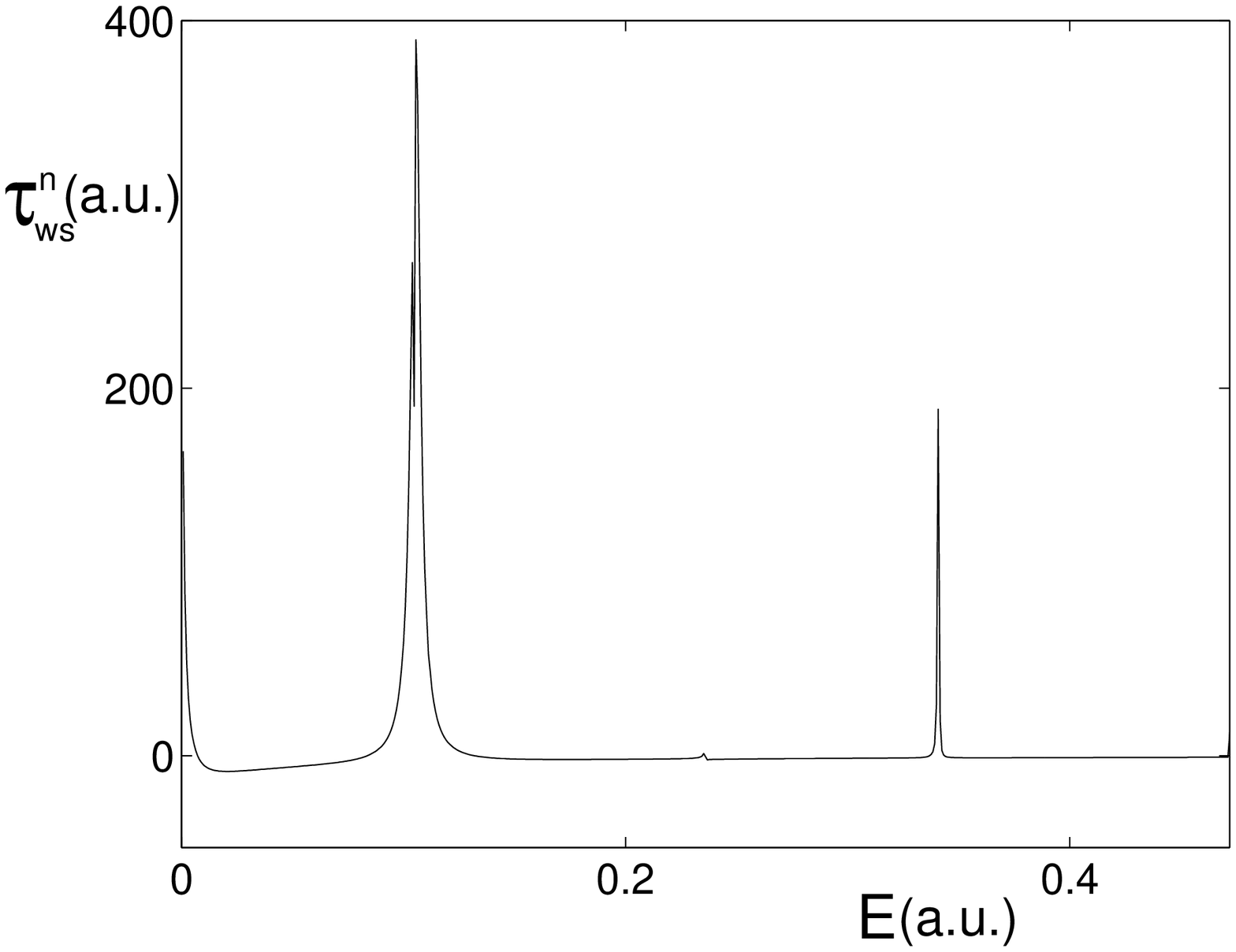}
\caption{}
\label{fig:wig05}
\end{centering}
\end{figure}

\begin{figure}
\begin{centering}
\leavevmode
\epsfxsize=0.45\linewidth
\epsfbox{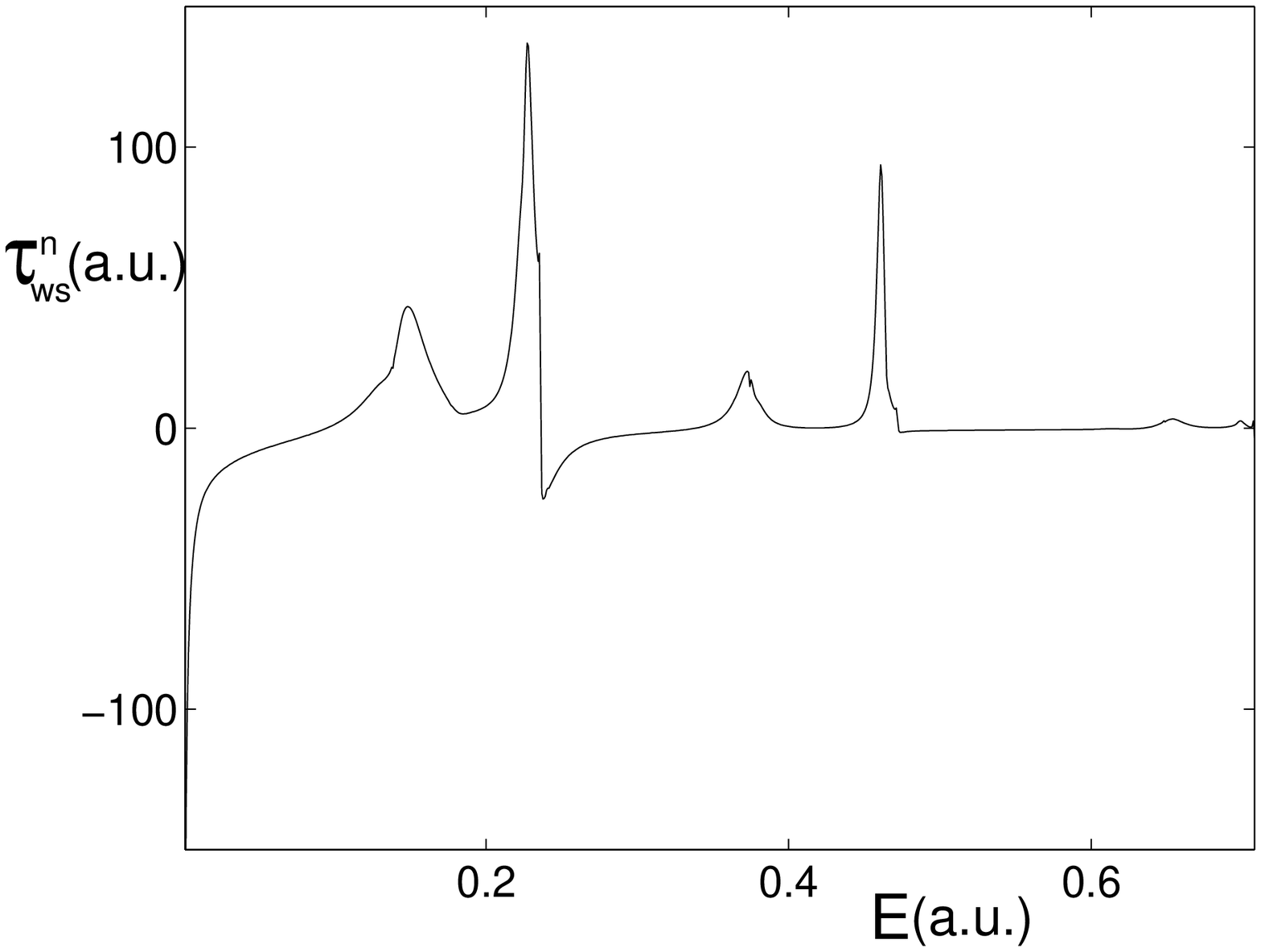}
\caption{}
\label{fig:wig225}
\end{centering}
\end{figure}

\begin{figure}
\begin{centering}
\leavevmode
\epsfxsize=0.5\linewidth
\epsfbox{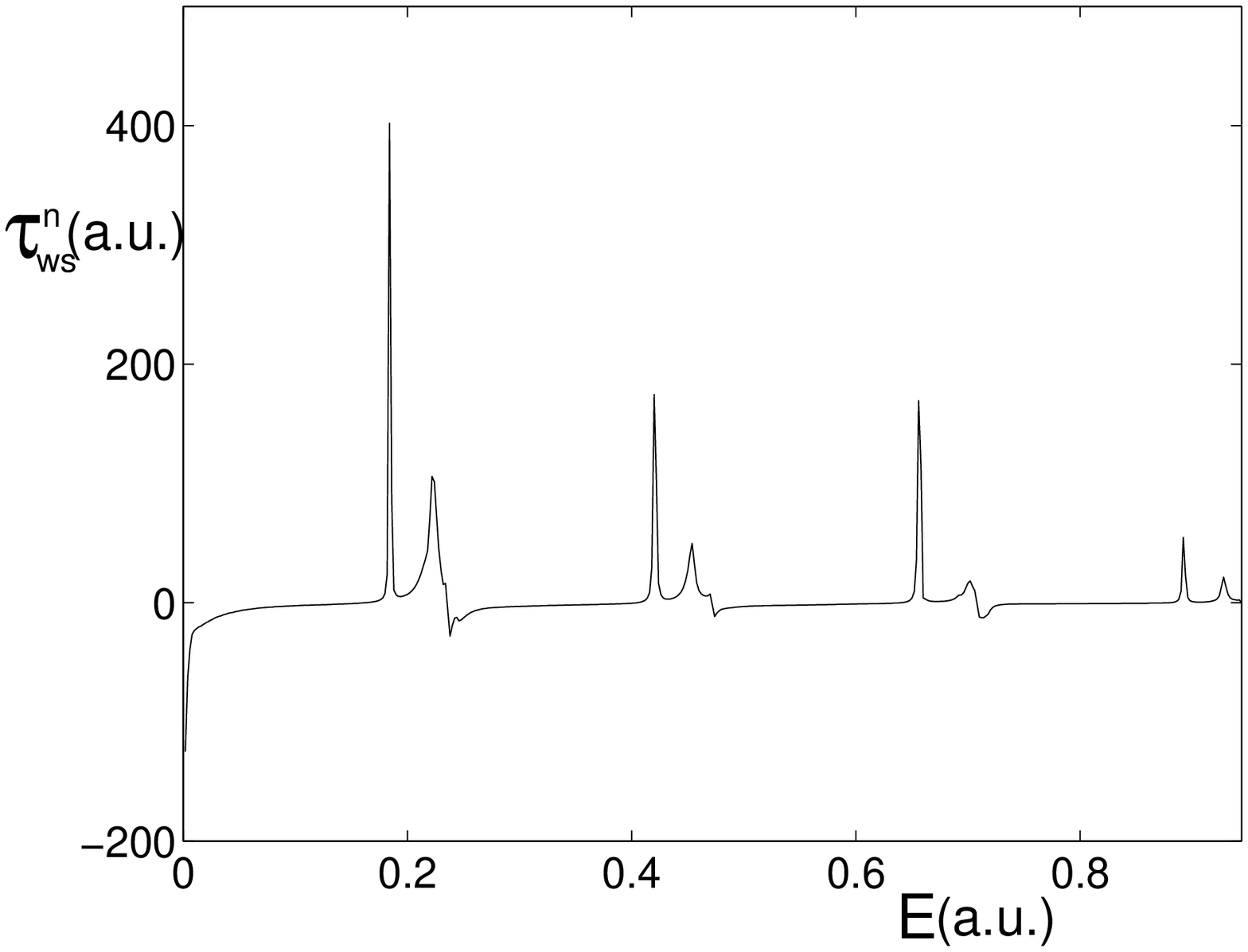}
\caption{}
\label{fig:wig525}
\end{centering}
\end{figure}

\begin{figure}
\begin{centering}
\leavevmode
\epsfxsize=0.5\linewidth
\epsfbox{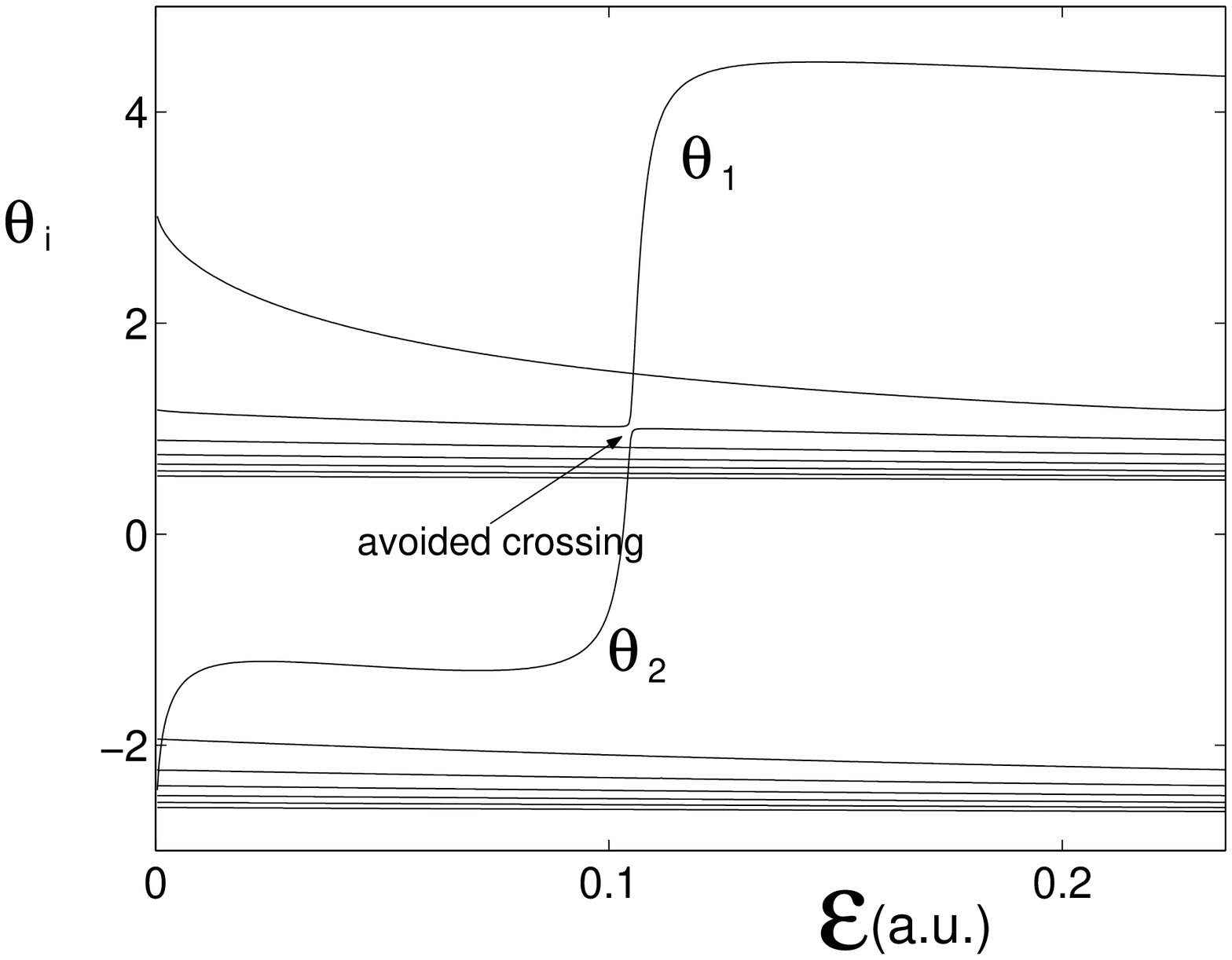}
\caption{}
\label{fig:phases05}
\end{centering}
\end{figure}

\newpage
\begin{center}
LIST OF FIGURES
\end{center}
\begin{enumerate}
\item Figure1: The fourier components $\frac{V_{n}(\alpha_{0};x)}{i^{n}}$ (a.u.)
of the inverted Gaussian potential as a function of the one space dimension $x$
(a.u.) in the K-H frame, for $\alpha_{0}=2.25$ a.u..

\item Figure2: Not drawn to scale, are shown in the K-H frame the asymptotic 
regions I $x\in [x_{0},+\infty)$ (a.u.) and III $x\in (-\infty,-x_{0}]$ (a.u.), where the potential is asymptotically zero, and the scattering region II where the 
inverted Gaussian potential oscillates laterally. In regions I and III, we 
also show the Floquet channels, denoted by dotted lines, and the incoming and 
outgoing electron waves, denoted by solid arrows.

\item Figure3: The transmission coefficients $\it{T_{n}}$ and $\it{T_{tot,n}}$,
 respectively, as a function of electron incident energy $E$, with $E\in [0,2\omega)$ a.u., for $\alpha_{0}=0.5$ a.u.. There is only one Fano transmission 
resonance at $E=0.106$ a.u., associated with the first quasibound
state.

\item Figure4: The transmission coefficients $\it{T_{n}}$ and $\it{T_{tot,n}}$,
 respectively, as a function of electron incident energy $E$, with $E\in [0,2\omega)$ a.u., for $\alpha_{0}=2.25$ a.u..
 There are two Fano transmission resonances at $E=0.142$ a.u. and $E=0.225$ 
a.u., associated with the first and second quasibound states,
  respectively.
 The second order Fano transmission resonances for $E > \omega$ are more
prominent than those for $\alpha_{0}=0.5$ a.u..

\item Figure5: The transmission coefficients $\it{T_{n}}$ and $\it{T_{tot,n}}$
as a function of electron incident energy $E$, with $E\in [0,2\omega)$ a.u., for $\alpha_{0}=5.25$ a.u.. There are two Fano transmission resonances at 
$E=0.185$ a.u. and $E=0.219$ a.u., associated with the first and 
second quasibound states, respectively. The second order Fano transmission 
resonances for $E > \omega$ are more prominent than those for
 $\alpha_{0}=2.25$ a.u..

\item Figure6: Contour plot of the transmission coefficient $\it{T_{n}}$
in the complex energy plane for $\alpha_{0}=2.25$ a.u.. The 
dark$\rightarrow$light areas correspond to increasing values of $\it{T_{n}}$. 
There are two zero-pole pairs each associated with the Fano resonances in 
Fig.(\ref{fig:T225}). From the poles
we determine the real part and the lifetime, $\tau_{L}$, of the first and second 
quasibound states.  

\item Figure7: Real part of the first (squares) and second (dots) quasibound states minus a photon energy as a function of $\alpha_{0}$ (a.u.). The real part of the quasibound states
is found from the poles of the transmission coefficient $\it{T_{n}}$ in the complex
energy plane. 
 
\item Figure8: Ionization rate of the first (squares) and second (dots) 
quasibound states as a function of $\alpha_{0}$ (a.u.). The imaginary part of the 
quasibound states is found from the poles of the transmission coefficient 
$\it{T_{n}}$ in the complex energy plane. 

\item Figure9: The Wigner-Smith delay times, $\tau_{ws}^{n}$, as a function of electron 
incident energy $E\in[0,2\omega)$ for $\alpha_{0}=0.5$ a.u.. There is one peak at $E=0.106$ a.u. and smaller peaks at higher order resonances, 
 associated with the Fano resonance in Fig.(\ref{fig:T05}). For small incident energy, $E$, the 
Wigner-Smith delay time is positive.  

\item Figure10: The Wigner-Smith delay times, $\tau_{ws}^{n}$, as a function of 
electron incident energy $E\in[0,3\omega)$ for $\alpha_{0}=2.25$ a.u.. There are two peaks
at $E=0.142$ a.u. and $E=0.225$ a.u. and smaller peaks at higher order resonances, associated with the two Fano resonances at
Fig.(\ref{fig:T225}). For small incident energy, $E$, the Wigner-Smith delay time is negative.  

\item Figure11: The Wigner-Smith delay times, $\tau_{ws}^{n}$, as a function of 
electron incident energy $E\in[0,4\omega)$ for $\alpha_{0}=5.25$ a.u.. There are two peaks
at $E=0.185$ a.u. and $E=0.219$ a.u. and smaller peaks at higher order resonances, associated with the two Fano resonances at
Fig.(\ref{fig:T525}). For small incident energy, $E$, the Wigner-Smith delay time is negative.  

\item Figure12: The eigenphases $\theta_{i}$ (rad) as a function of Floquet
energy $\mathcal{E}$ (a.u.) for $\alpha_{0}=0.5$ a.u.. 
 For $a_{0}\ll 0.5 $ a.u. the eigenphases $\theta_{1}$ and $\theta_{2}$ 
intersect each other as a function
of Floquet energy $\mathcal{E}$ (a.u.). It is only as $\alpha_{0}$ is increased 
that the eigenphases repel as a function of $\mathcal{E}$ (a.u.) and form an ``avoided
crossing'', indicated by an arrow.   

\item Figure13: For $\alpha_{0}=0.5$ we retain $n=-6,...,0,...6$ channels
and obtain $14$ eigenphases from the $14\times 14$ Floquet 
$\mbox{\boldmath$S$}$-matrix. Only two eigenphases $\theta_{1}$ and $\theta_{2}$
 participate in the ``avoided crossing''. The eigenphases $\theta_{1}$ 
and $\theta_{2}$ 
exchange character completely at the sharp ``avoided crossing'' shown in 
Fig.(\ref{fig:phases05}). To 
show, quantitatively, how the character exchange takes place we plot in a) the
occupation probabilities $P_{0,1}$, $P_{1,1}$ and $P_{2,1}$ of the $|\theta_{1}>$ 
eigenvector on the propagating channels $n=0,1,2$ and in b) the
occupation probabilities $P_{0,2}$, $P_{1,2}$ and $P_{2,2}$ of the $|\theta_{2}>$ 
eigenvector on the propagating channels $n=0,1,2$ as a function of 
Floquet energy $\mathcal{E}$ (a.u.). Before the avoided crossing the eigenvector $|\theta_{1}>$ 
has support on channel $n=1$ and the eigenvector $|\theta_{2}>$ mainly on channel $n=0$, while 
after the avoided crossing the eigenvector $|\theta_{1}>$ 
has support mainly on channel $n=0$ and the eigenvector $|\theta_{2}>$ on channel $n=1$, thus exchanging character completely. 

\item Figure14: The eigenphases $\theta_{i}$ (rad) as a function of Floquet
energy $\mathcal{E}$ (a.u.) for $\alpha_{0}=1.25$ a.u.. The eigenphases 
$\theta_{1},...,\theta_{5}$ 
participate in the ``avoided crossings'' shown in Table II. 

\item Figure15: The eigenphases $\theta_{i}$ (rad) as a function of 
Floquet energy $\mathcal{E}$ (a.u.) for $\alpha_{0}=2.25$ a.u.. The eigenphases 
$\theta_{1},...,\theta_{7}$ participate in the ``avoided crossings'' shown in Table III. 

\item Figure16: Strobe plots of the classical dynamics, for the inverted Gaussian
in the presence of the driving field, in the Lab frame for a) $\alpha_{0}=0.5$,
 b) $\alpha_{0}=1.25$ and c) $\alpha_{0}=2.25$.
The initial conditions used to generate the plots lie on the line $p=0$ as well
as on the lines with $-1<p<1$. The location of the period-1 orbits are indicated by filled
squares. The period-1 orbits are located at a) $-1.87$, $-0.36$ and $2.46$ b) $2.84$
and c) $1.88$, $3.38$ and $4.54$. For very small values of the driving field $\alpha_{0}$ (not shown)
there is a large regular island around the region at $x=0$, $p=0$.
As $\alpha_{0}$ is increased to $0.5$ there are two regular islands reduced in size
indicating the destruction of the KAM tori. As $\alpha_{0}$ is further increased to $1.25$
and $2.25$ the regular islands disappear and the phase space is dominated by chaos.

\item Figure 17: Strobe plots of the classical dynamics, for the inverted Gaussian
in the presence of the driving field, in the Lab frame for a) $\alpha_{0}=0.5$
and b) $\alpha_{0}=2.25$. The initial conditions of the classical
momenta used to generate the plots are chosen to correspond to the middle of the 
Floquet propagating channels, that is the initial conditions lie on the lines 
$p=\pm \sqrt{2(0.13+n\omega)}$ 
with $n=0,1,...,8$. As the strength of the driving field is increased more trajectories
get pulled in the chaotic region in the classical phase space.

\end{enumerate}
 
\newpage

\begin{figure}
\begin{centering}
\leavevmode
\epsfxsize=0.5\linewidth
\epsfbox{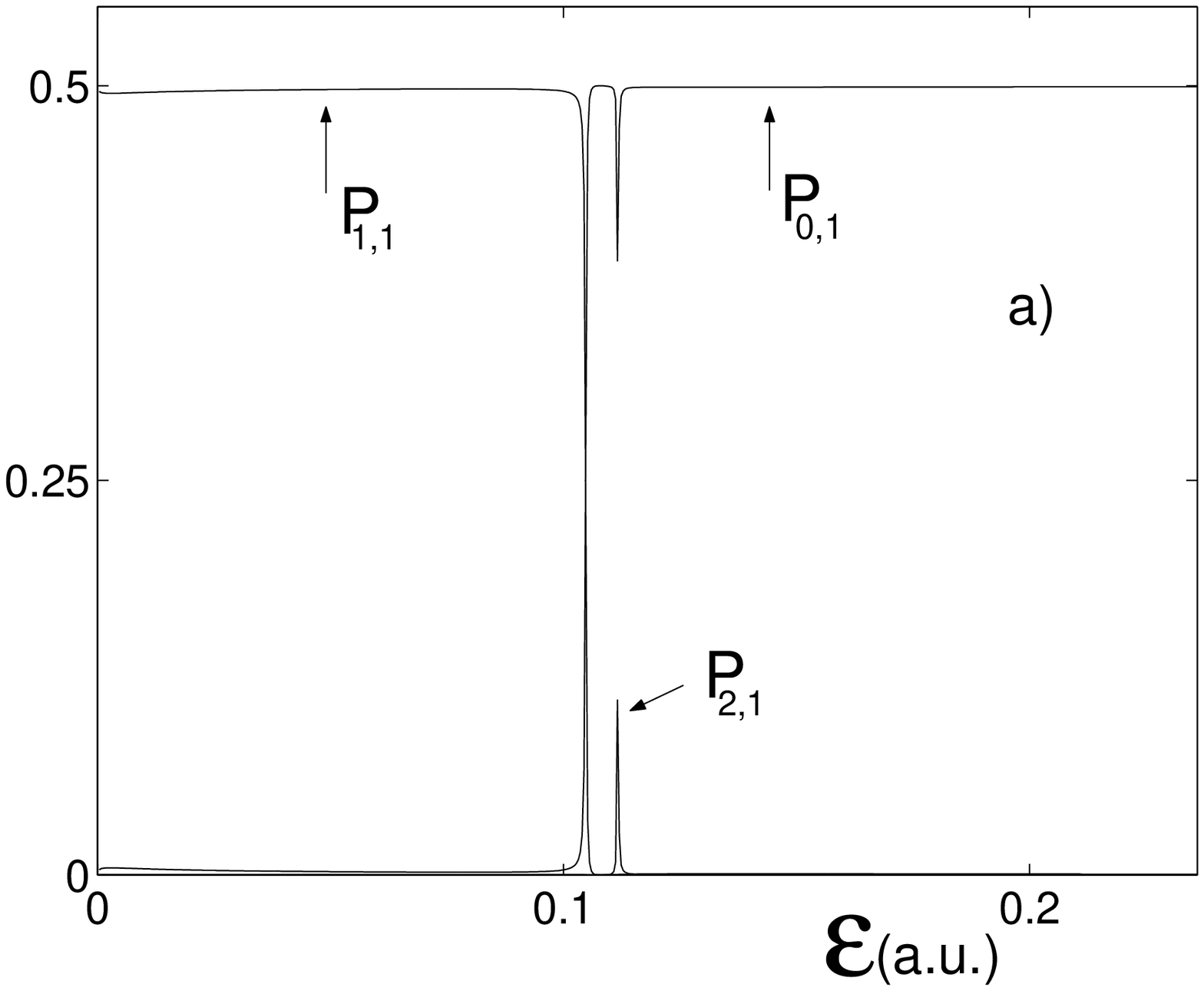}
\leavevmode
\epsfxsize=0.5\linewidth
\epsfbox{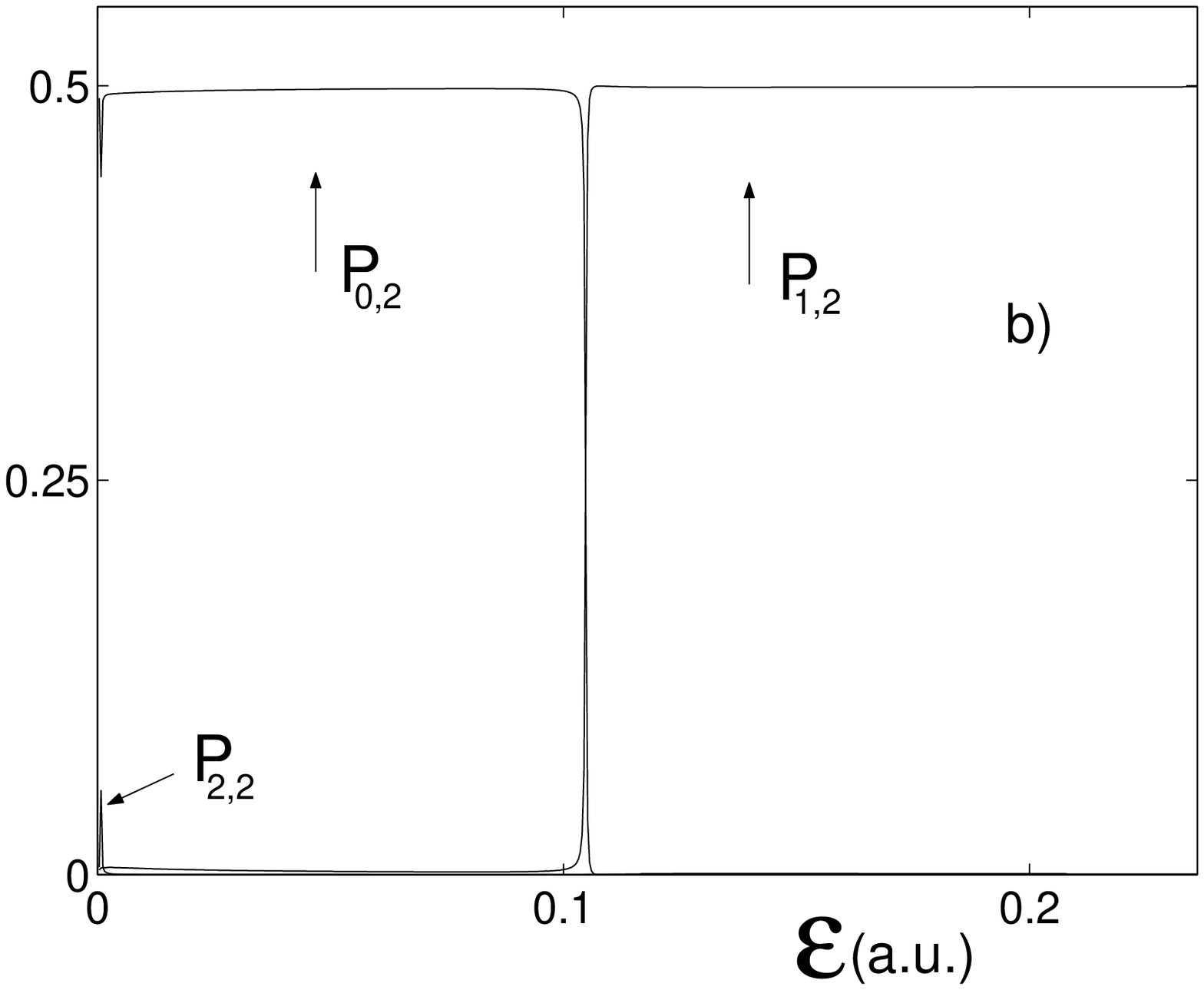}
\caption{}
\label{fig:prob05}
\end{centering}
\end{figure}

\begin{figure}
\begin{centering}
\leavevmode
\epsfxsize=0.5\linewidth
\epsfbox{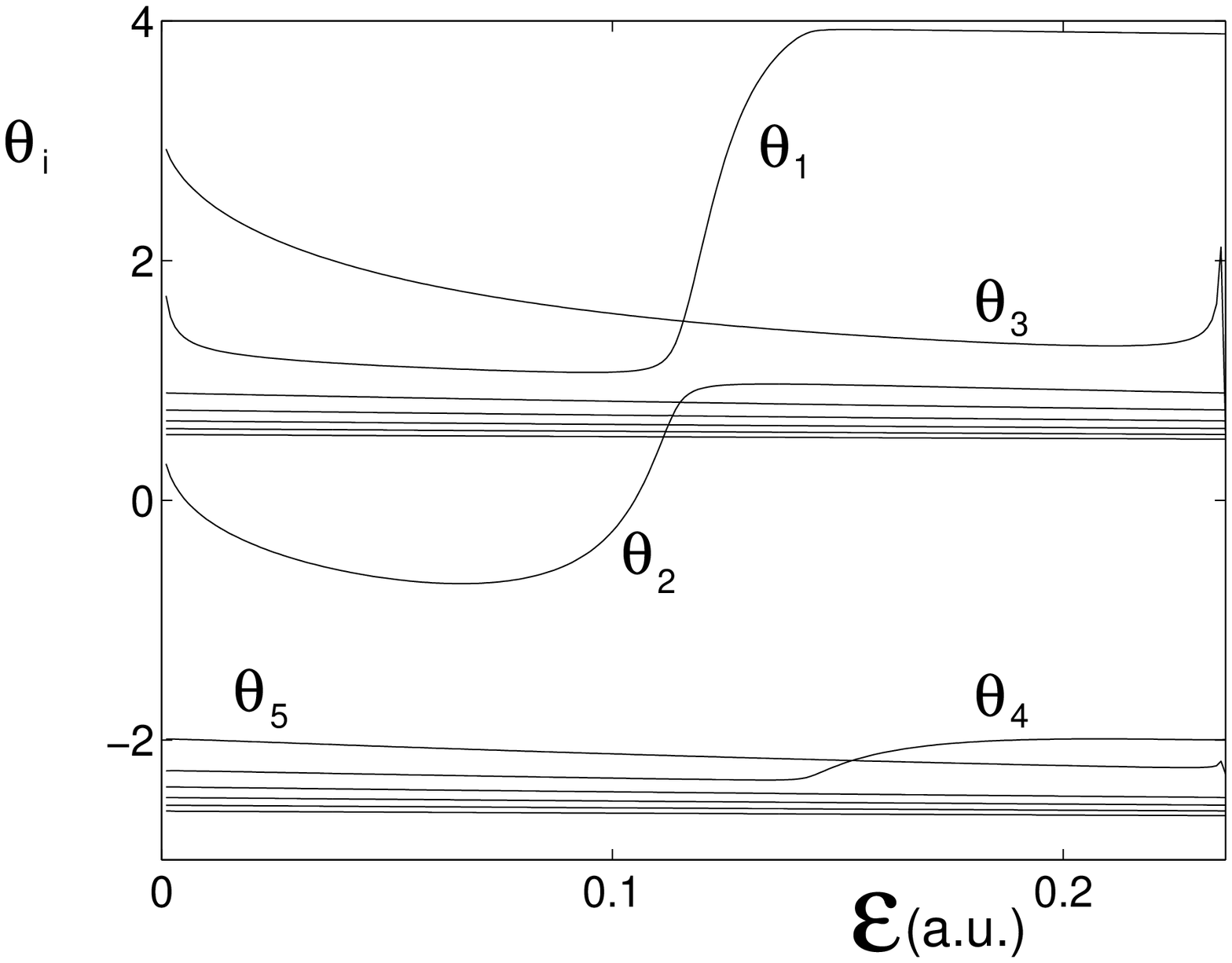}
\caption{}
\label{fig:phases125}
\end{centering}
\end{figure}

\begin{figure}
\begin{centering}
\leavevmode
\epsfxsize=0.50\linewidth
\epsfbox{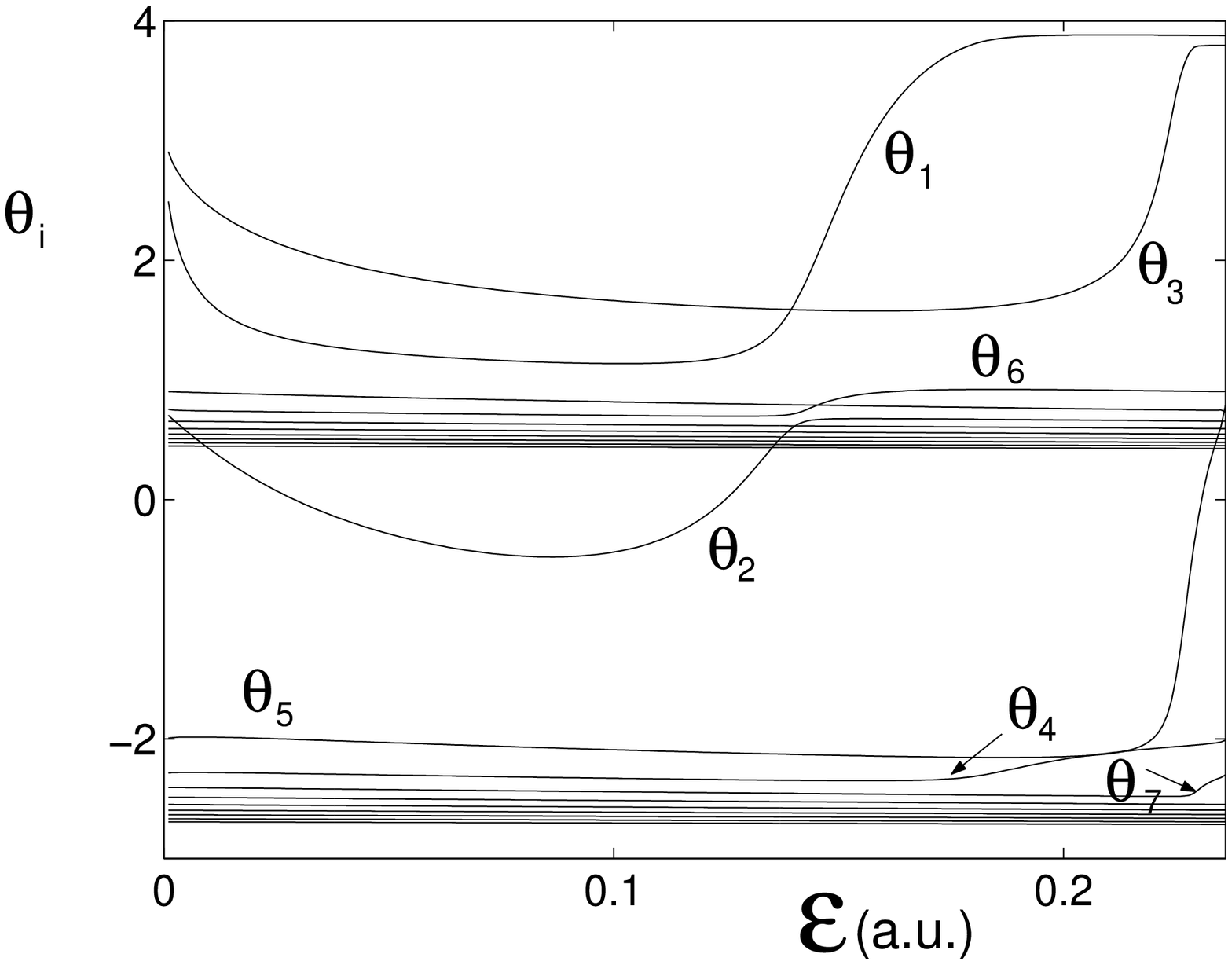}
\caption{}
\label{fig:phases225}
\end{centering}
\end{figure}

\begin{figure}
\begin{centering}
\leavevmode
\epsfxsize=0.5\linewidth
\epsfbox{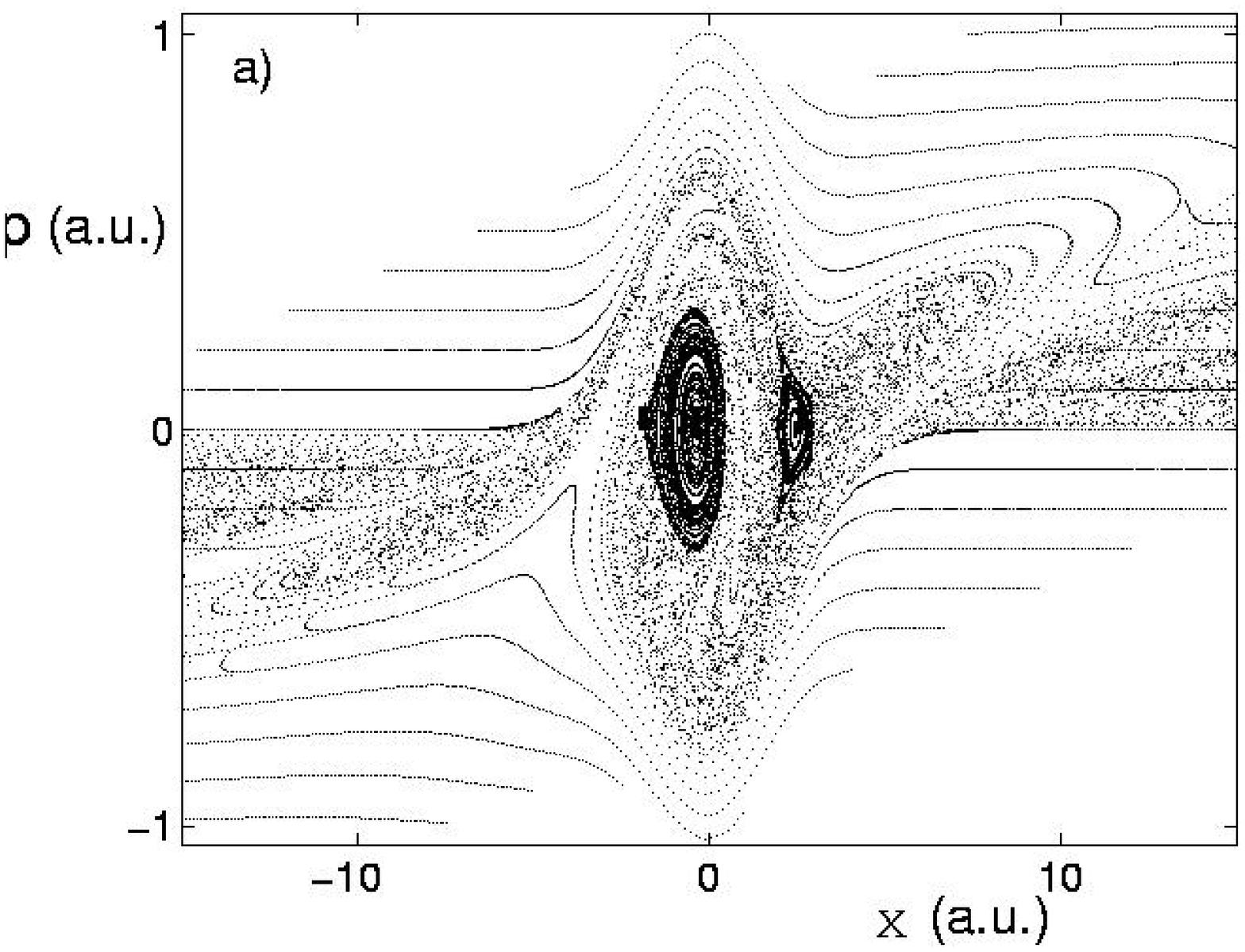}
\end{centering}
\begin{centering}
\leavevmode
\epsfxsize=0.5\linewidth
\epsfbox{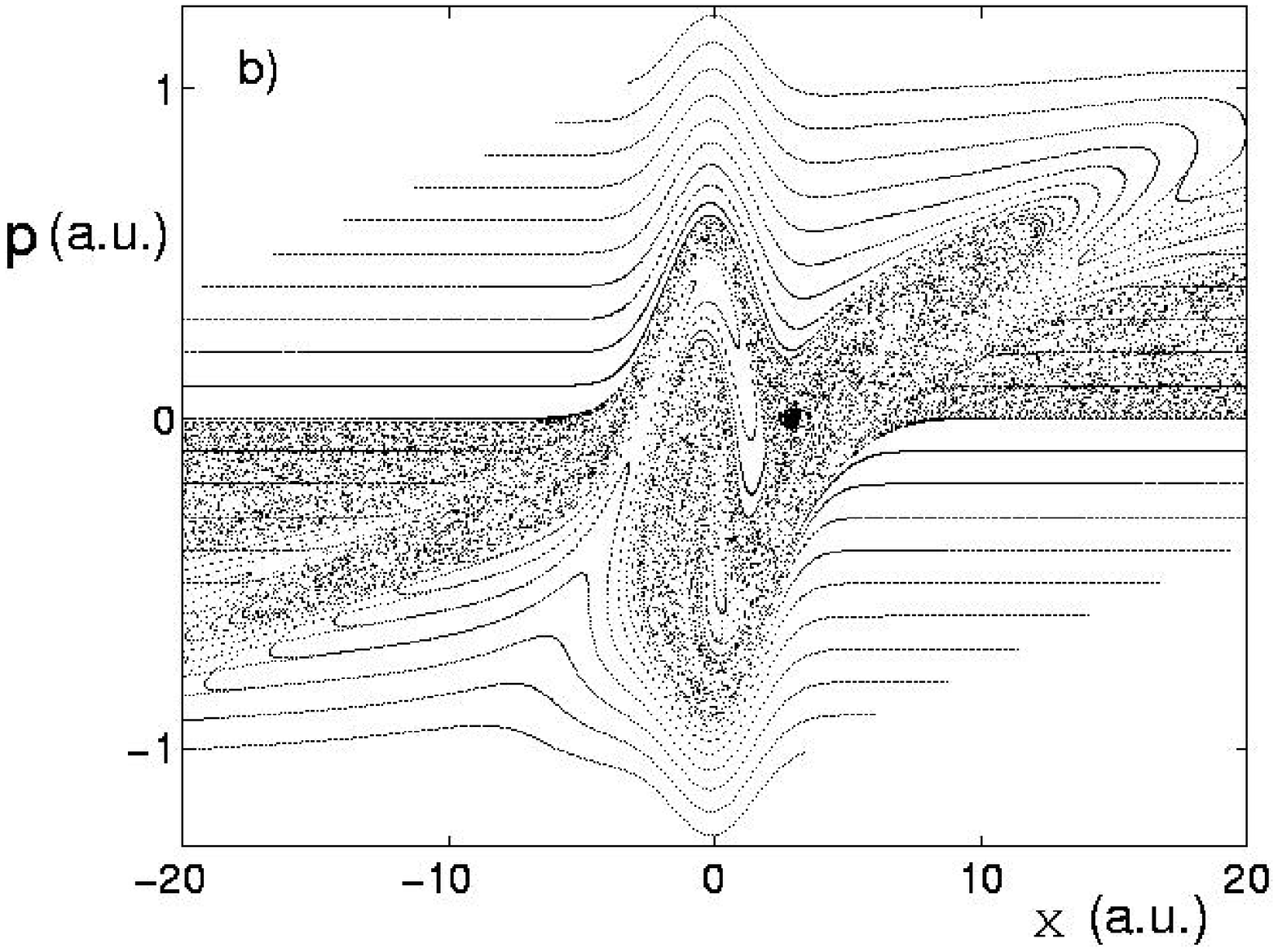}
\end{centering}
\begin{centering}
\leavevmode
\epsfxsize=0.5\linewidth
\epsfbox{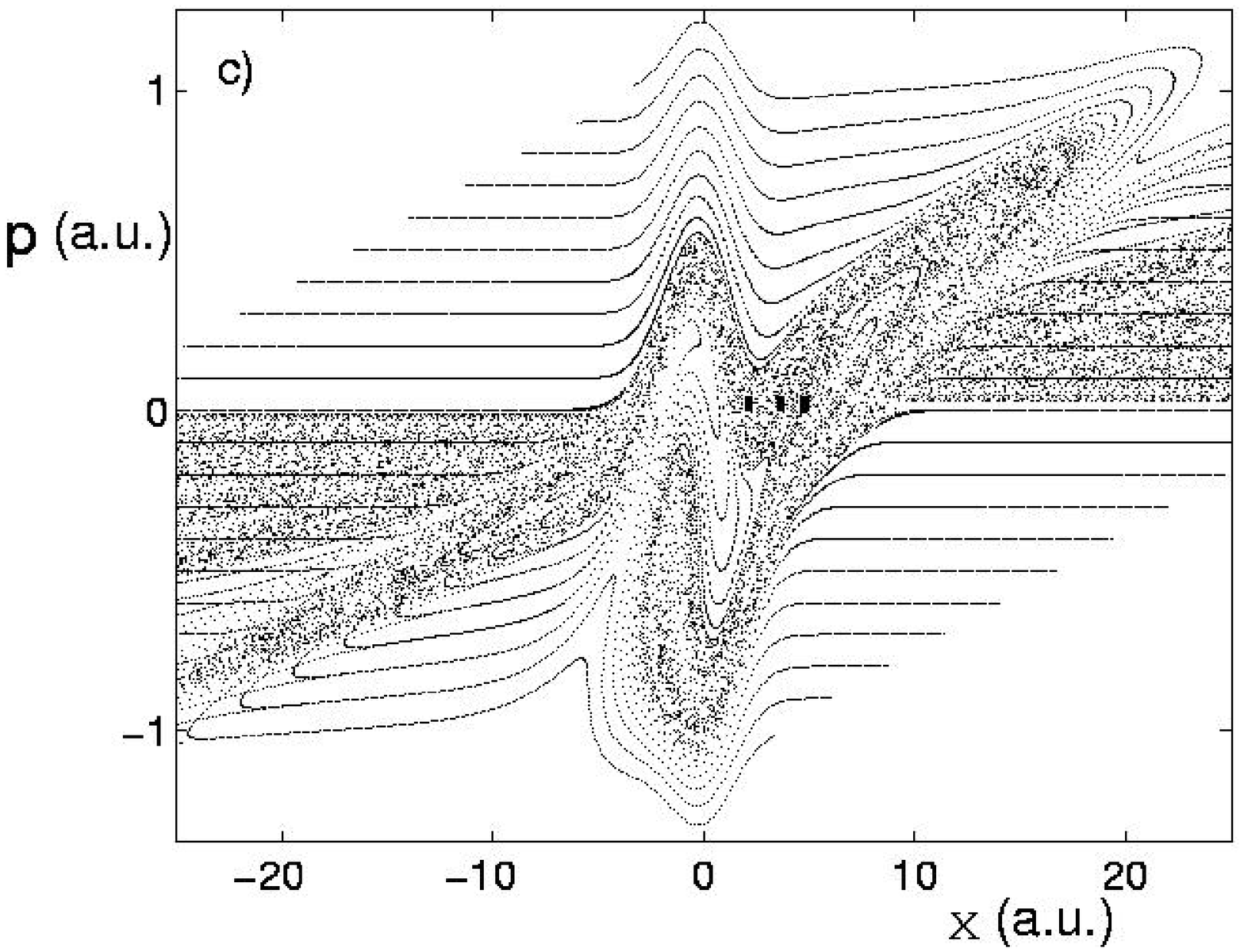}
\caption{}
\label{fig:px}
\end{centering}
\end{figure}

\begin{figure}
\begin{centering}
\leavevmode
\epsfxsize=0.75\linewidth
\epsfbox{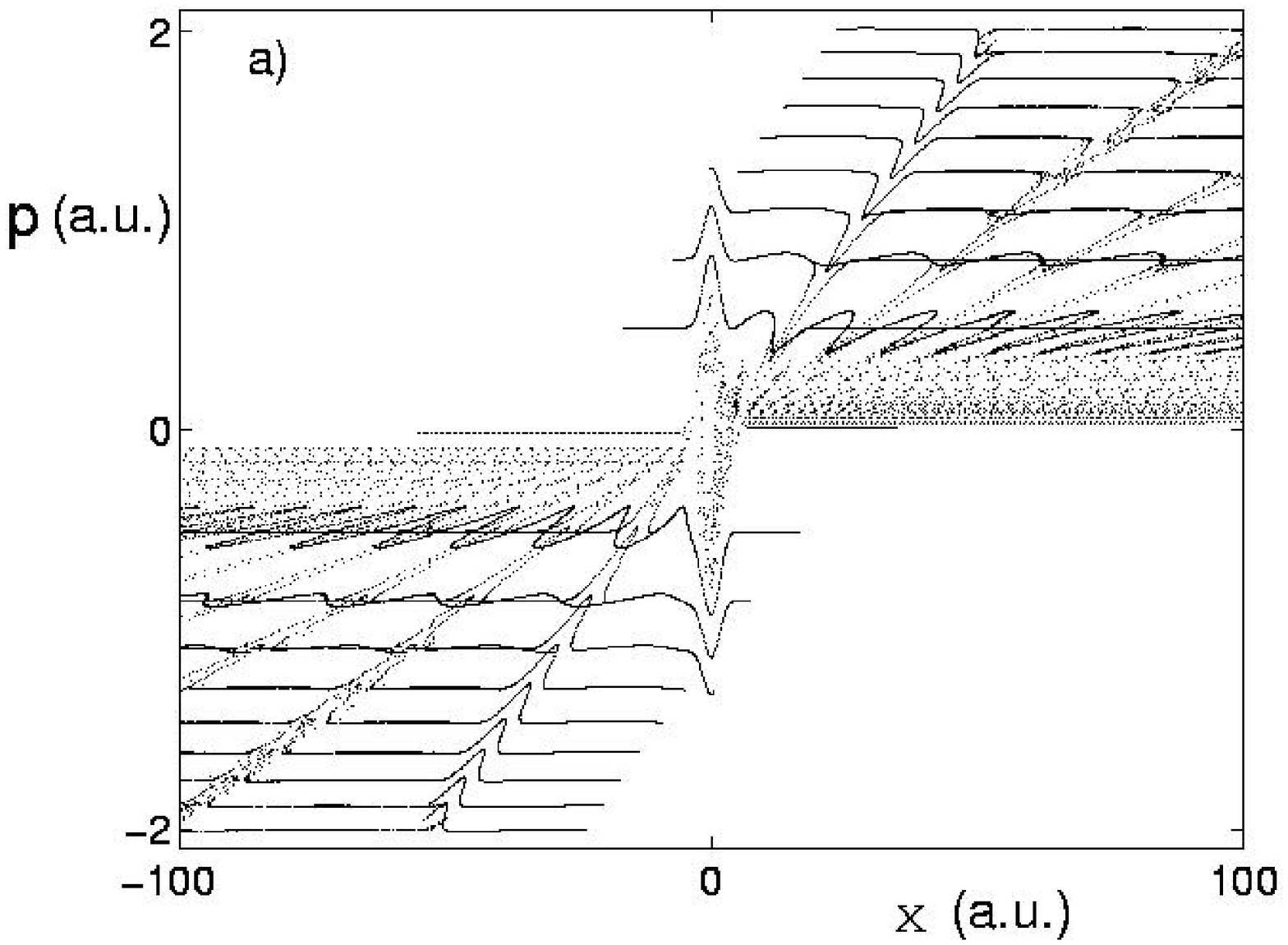}
\end{centering}
\begin{centering}
\leavevmode
\epsfxsize=0.75\linewidth
\epsfbox{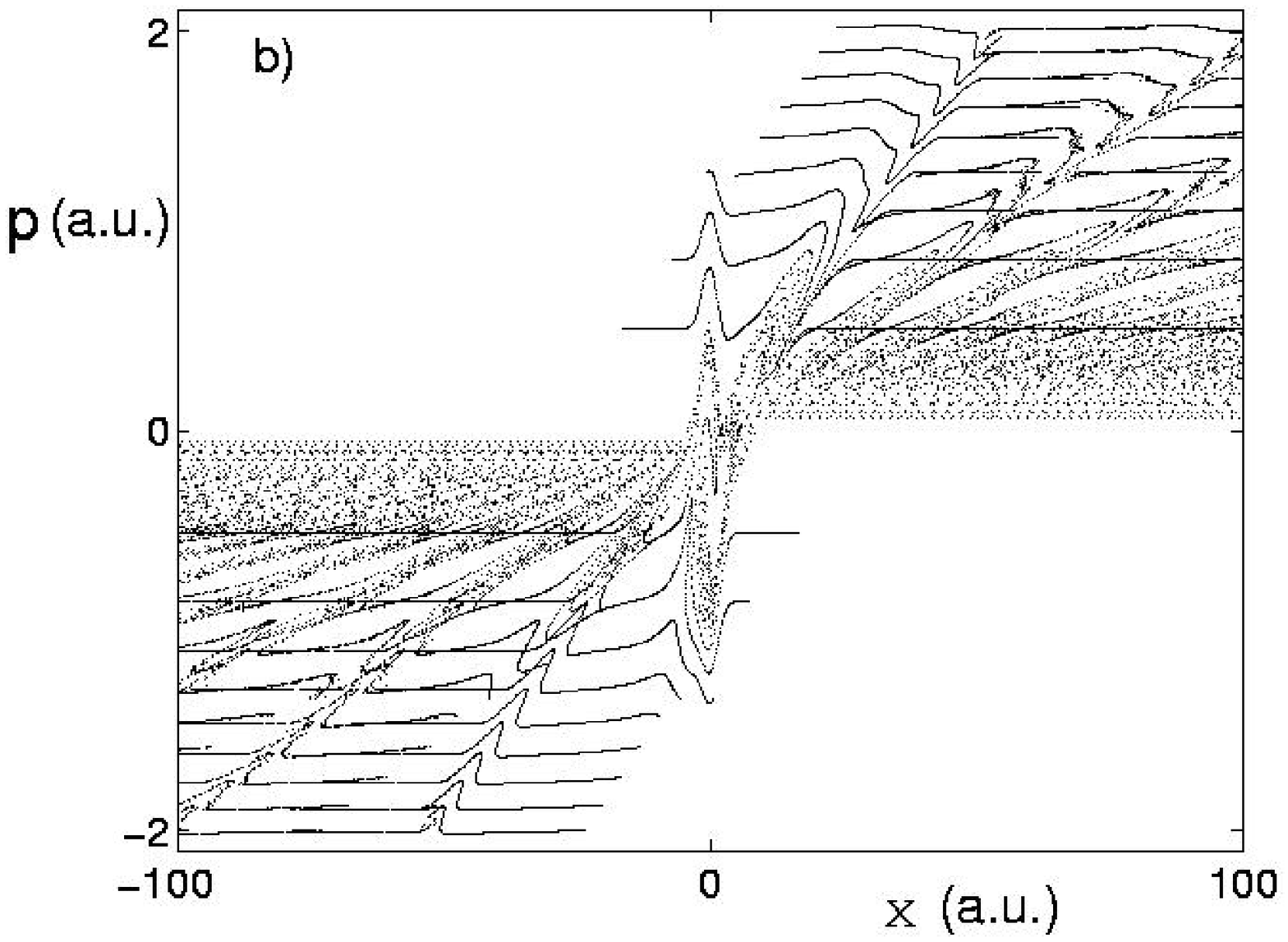}
\caption{}
\label{fig:pt}
\end{centering}
\end{figure}


\begin{references}
\bibitem{ion1}
M. Gavrila and J. Z. Kaminski, Phys. Rev. Lett. 52, 613 (1984)
\bibitem{ion2} M. Pont, N. R.
 Walet, M. Gavrila, and C. W. McCurdy, Phys. Rev. Lett. 61, 939 (1988); M. Pont and
M. Gavrila, Phys. Rev. Lett. 65, 2362 (1990)
\bibitem{ion3}
K. Burnett, P. L. Knight, B. R. M. Piraux, and V. C. Reed, Phys. Rev. Lett 66, 301 (1991);
K. C. Kulander, K. J. Schafer, and J. L. Krause, Phys. Rev. Lett. 66, 2601 (1991);
Q. Su, J. H. Eberly, and J. Javanainen, Phys. Rev. Lett. 64, 862 (1990)  
\bibitem{experim1}
M. P. de Boer, J. H. Hoogenraad, R. B. Vrijen, R. C. Constantinescu, L. D. Noordam, and 
H. G. Muller, Phys. Rev. A 50, 4085 (1994); N. J. van Druten, R. C. Constantinescu, J. M.
Schins, H. Nieuwenhuize and H. G. Muller, Phys. Rev. A 55, 622 (1997)
\bibitem{experim2}
C. O. Reinhold, J. Burgd\mbox{\"{o}}rfer, M. T. Frey, and F. B. Dunning, Phys. Rev. Lett. 79, 5226 (1997)
\bibitem{Kramers}
H. A. Kramers, {\it{Collected Scientific Papers}} (North-Holland, Amsterdam, 1956), 
 p.272.
\bibitem{Henneberger}
W. C. Henneberger, Phys. Rev. Lett. 21, 838 (1968).
\bibitem{Dimou}
L. Dimou and F. H. M. Faisal, Phys. Rev. Lett. 59, 872 (1987);
 L. A. Collins and G. Csanak, Phys. Rev. A 44, R5343 (1991) 
\bibitem{Bhatt}
R. Bhatt, B. Piraux, and K. Burnett, Phys. Rev. A 37, 98 (1988) 
\bibitem{Bardsley}
J. N. Bardsley and M. J. Comella, Phys. Rev. A 39, 2252 (1989)
\bibitem{Yao}
G. Yao and S.-I Chu, Phys. Rev. A 45, 6735 (1992)
\bibitem{Marinescu}
M. Marinescu and M. Gavrila, Phys. Rev. A 53, 2513 (1996)
\bibitem{Timberlake2}
T. Timberlake and L. E. Reichl, to appear in Phys. Rev. A, vol.64 
\bibitem{Reichl1}
W. Li and L. E. Reichl, Phys. Rev. B 60, 15732 (1999) 
\bibitem{Floquet}
J. H. Shirley, Phys. Rev. 138, B979 (1965)
\bibitem{Fearnside}
A. S. Fearnside, R. M. Portvliege, and R. Shakeshaft, Phys. Rev. A 51, 1471 (1995)
\bibitem{Fano}
U. Fano, Phys. Rev. 124, 1866 (1961)
\bibitem{Tekman}
E. Tekman and P. F. Bagwell, Phys. Rev. B 48, 2553 (1993)
\bibitem{pairs}
W. Porod, Zhi-an Shao, and C. S. Lent, Phys. Rev. B 48, 8495 (1993)
\bibitem{Wigner}
E. P. Wigner, Phys. Rev. 98, 145 (1955)
\bibitem{Smith}
F. T. Smith, Phys. Rev. 118, 349 (1960)
\bibitem{Na}
K. Na and L. E. Reichl, Phys. Rev. B 59, 13073 (1999)
\bibitem{bound}
T. Timberlake and L. E. Reichl, Phys. Rev. A 59, 2886 (1999)
\bibitem{Henseler}
M. Henseler, T. Dittrich, and K. Richter (submitted to Phys. Rev. E)
\bibitem{Numerov}
{\it{Methods in Computational Physics}} edited by B. Alder et. al. 
 (Academic Press New York and London 1966), p.16 
\end{references}
\end{document}